\documentclass[twocolumn,useamsfonts]{pasj02} 
\usepackage{url}
\usepackage[switch,mathlines]{lineno} 
\usepackage{natbib}
\usepackage{bm}
\usepackage{comment}
\usepackage[dvipsnames]{xcolor}
\usepackage{xspace}
\usepackage{graphicx}
\usepackage{url}
\usepackage{amsmath}
\usepackage{amssymb}

\newcommand{\SNR}{\mathrm{SNR}}
\usepackage{ulem} 

\jyear{2026}
\Received{}
\Accepted{}

\begin{document} 

\title{Denoising weak lensing mass maps with diffusion model: systematic comparison with generative adversarial network}

\author{
 Shohei~\textsc{D. Aoyama},\altaffilmark{1}\altemailmark
 Ken~\textsc{Osato},\altaffilmark{2,1,3,4}\altemailmark\orcid{0000-0002-7934-2569}
 and
 Masato~\textsc{Shirasaki}\altaffilmark{5,6,3}\altemailmark\orcid{0000-0002-1706-5797}
}
\altaffiltext{1}{Department of Physics, Graduate School of Science, Chiba University, Chiba 263-8522, Japan}
\altaffiltext{2}{Center for Frontier Science, Chiba University, Chiba 263-8522, Japan}
\altaffiltext{3}{RIKEN Center for Advanced Intelligence Project, Chuo, Tokyo 103-0027, Japan}
\altaffiltext{4}{Kavli Institute for the Physics and Mathematics of the Universe,
                 The University of Tokyo Institutes for Advanced Study,
                 Chiba, 277-8583, Japan}
\altaffiltext{5}{National Astronomical Observatory of Japan (NAOJ),
                 National Institutes of Natural Science,
                 Mitaka, Tokyo 181-8588, Japan}
\altaffiltext{6}{The Institute of Statistical Mathematics, Tachikawa, Tokyo 190-8562, Japan}
\email{shohei.aoyama@chiba-u.jp,ken.osato@chiba-u.jp,masato.shirasaki@nao.ac.jp}



\maketitle

\begin{abstract}
Weak gravitational lensing (WL) is the unique and powerful probe into the large-scale structures of the Universe. Removing the shape noise from the observed WL field, i.e., denoising, enhances the potential of WL by accessing information at small scales where the shape noise dominates without denoising. We utilise two machine learning (ML) models for denosing: generative adversarial network (GAN) and diffusion model (DM). We evaluate the performance of denosing with GAN and DM utilising the large suite of mock WL observations, which serve as the training and test data sets. We apply denoising to 1,000 noisy mass maps with GAN and DM models trained with 39,000 mock observations. Both models can fairly well reproduce the true convergence map on large scales. Then, we measure cosmological statistics: power spectrum, bispectrum, one-point probability distribution function, peak and minima counts, and scattering transform coefficients. We find that DM outperforms GAN in almost all considered statistics and recovers the correct statistics down to small scales. For example, the angular power spectrum can be recovered with DM up to multipoles $\ell \lesssim 6000$ while the noise power spectrum dominates from $\ell \simeq 2000$. We also conduct stress tests on the trained model; denoising the maps with different characteristics, e.g., different source redshifts, from the training data. The performance degrades at small scales, but the statistics can still be recovered at large scales. Though the training of DM is more computationally demanding compared with GAN, there are several advantages: numerically stable training, higher performance in the reconstruction of cosmological statistics, and sampling multiple realisations once the model is trained. It has been known that DM can generate higher-quality images in real-world problems than GAN, the superiority has been confirmed as well in the WL denoising problem.
\end{abstract}


\section{Introduction}
\label{sec:introduction}
Weak gravitational lensing (WL) is one of the major probes into the matter distribution in the Universe.
Through the statistical analysis of the matter distribution mapped with WL,
one can address fundamental physics such as the nature of dark matter and dark energy
\citep[for a review, see, e.g.,][]{Weinberg2013}.
Stage-III WL surveys, including Dark Energy Survey \citep[DES;][]{DES2016},
Subaru Hyper-Suprime Cam \citep[HSC;][]{Aihara2018,Dalal2023,Li2023},
and Kilo Degree Survey \citep[KiDS;][]{Hildebrandt2017,Heymans2021,Wright2025},
have already proven the capability of WL to constrain cosmological parameters precisely.
Furthermore, Stage-IV WL surveys such as \textit{Euclid} \citep{Euclid},
Vera C. Rubin Observatory Legacy Survey of Space and Time \citep[LSST;][]{LSST}
and \textit{Nancy Grace Roman Space Telescope} \citep{WFIRST} will enable the WL measurement with unprecedented precision
due to wider area coverage and deeper imaging.

WL refers to small distortions of shapes of distant galaxies, which are sensitive to the foreground gravitational potential.
Through the measurement of the galaxy shapes, one can reconstruct the projected density field.
The analysis is referred to as mass reconstruction \citep{Kaiser1993,Schneider1995,Bartelmann1995}
and the resultant two-dimensional projected density field is called \textit{mass map}.
In mass reconstruction analysis, the number of source galaxies is finite, and thus,
the intrinsic shape of source galaxies dilutes the WL signal due to the Poissonian nature.
This noise is referred to as shape noise and is the major contaminant in mass maps.
The statistical property of the shape noise is well studied \citep[e.g.,][]{vanWaerbeke2000},
and various approaches to reduce the shape noise have been proposed \citep{Bartelmann1996,Bridle1998,Marshall2002,Starck2006,Deriaz2012}.
Precise map-level reduction of shape noise, which is referred to as \textit{denoising}, is critical
for cluster search \citep{Hamana2004,Shirasaki2015} and identification of filamentary structures \citep{Higuchi2014}.

Recently, machine learning (ML) has emerged as a promising approach to accurate and fast mass reconstruction \citep[e.g.,][]{Jeffrey2020}.
In particular, generative models in the image domain, which take two-dimensional images as inputs and outputs,
are suitable for denoising mass maps.
For example, the generative adversarial network \citep[GAN;][]{Goodfellow2014} trained with mock WL simulations
demonstrate high performance in denoising mass maps \citep{Shirasaki2019,Whitney2025} and
surrogating gravitational evolution for WL \citep{Yiu2022,Shirasaki2024,Boruah2024}.
In this work, we focus on the diffusion model \citep[DM;][]{Sohl-Dickstein2015},
which is one of the most powerful generative models.
DM is also classified as a deep generative model and
consists of two processes: the forward process, which adds noise to the input field,
and the reverse process, which removes the noise from the noisy data.
The forward process is analytically tractable, but the reverse process is not trivial,
and deep neural networks are employed to model the reverse process.
Starting from the white noise, the target data is reconstructed with iterations of the reverse process.
DM can handle various tasks relevant to image recognition, such as inpainting, super-resolution, and colourisation,
and demonstrates higher performance compared with existing generative models \citep{Dhariwal2021}.
In addition, DM has also been applied in mass reconstruction \citep{Remy2023,Boruah2025}.

We explore the capability of DM in denoising mass maps and conduct a systematic comparison with GAN.
We employ the large suite of mock WL simulations $\kappa$TNG \citep{Osato2021} to train the models.
$\kappa$TNG data set consists of 40,000 mass maps and we manually add the shape noise to simulate the realistic WL observations.
Finally, we construct 40,000 pairs of noiseless and noisy mass maps, and we use 39,000 maps for training and 1,000 maps for testing.
In contrast to GAN, DM can sample multiple realisations of denoised maps from a single noisy mass map.
For each noisy map, we sample five realisations with DM.
In order to mimic the multiple sampling for GAN, we train GAN networks five times with different weight initialisations.
Finally, there are 5,000 denoised maps both for GAN and DM.
Then, we compare the denoised mass maps at a pixel level and address the effect on the peaks in the mass maps.
Next, we measure the quantitative comparisons of cosmological statistics:
angular power spectrum, bispectrum, one-point probability distribution function, peak and minima counts, and scattering transform coefficients.
We also conduct stress tests, where denoising is applied to the modified test data, e.g., mass maps with source redshifts shifted,
to evaluate the robustness and flexibility of the trained models.

In Section~\ref{sec:WL_basics}, we present the basics of WL
and the shape noise in the mass map.
In Section~\ref{sec:methods}, we introduce the ML algorithms
to remove the shape noise from the mass map
and the data sets to train and test the algorithms.
In Section~\ref{sec:results}, we demonstrate how the ML approaches
denoise the mass map and the statistics can be reproduced.
In Section~\ref{sec:conclusions}, we make concluding remarks.

\section{Basics of weak lensing}
\label{sec:WL_basics}
Here, we review the basics of the formalism of gravitational lensing
\citep[for a review, see][]{Bartelmann2001,Kilbinger2015,Mandelbaum2018}.
Images of distant galaxies are distorted due to the gravitational potential
sourced from the intervening matter.
This gravitational lensing effect by the large-scale structures of the Universe
is referred to as \textit{cosmic shear}.
The distortion of galaxy shapes is characterised by the distortion matrix $A$:
\begin{equation}
    A_{ij} = \frac{\partial \beta_i}{\partial \theta_j}, \
    A =
    \begin{pmatrix}
        1 - \kappa - \gamma_1 & - \gamma_2 - \omega\\
        - \gamma_2 + \omega & 1 - \kappa + \gamma_1
    \end{pmatrix}
    ,
\end{equation}
where $\bm{\theta}$ is the observed angular position of a source object,
$\bm{\beta}$ is the true angular position, $\kappa$ is the convergence,
$\gamma_1$ and $\gamma_2$ are the shear, and $\omega$ is the rotation.
In the WL regime ($\phi/c^2 \ll 1$, where $\phi$ is the gravitational potential
and $c$ is the speed of light),
the convergence can be expressed as the projected density
convolved with the distance kernel $W_\kappa(\chi)$ along the line-of-sight direction: 
\begin{align}
    \kappa(\bm{\theta}) &= \int^{\chi_\mathrm{H}}_0 d\chi \, W_\kappa (\chi) \, \delta_\mathrm{m}(r(\chi)\bm{\theta}, \chi),\\
    \label{eq:kernel}
    W_\kappa (\chi) &= \frac{3 \Omega_\mathrm{m} H^2_0}{2c^2} r(\chi) \int^{\chi_\mathrm{H}}_\chi d\chi' \, p(\chi') \frac{r(\chi'-\chi)}{r(\chi')},
\end{align}
where $\chi$ is the comoving distance, $\chi_\mathrm{H}$ is the comoving distance to the horizon,
$\Omega_\mathrm{m}$ is the matter density at the present Universe,
$H_0$ is the Hubble constant,
$r(\chi)$ is the comoving angular diameter distance,
and $p(\chi)$ is the source galaxy distribution,
which is normalized as $\int d\chi \, p(\chi) = 1$.
Hereafter, since the flat Universe is assumed,
the comoving angular diameter distance is identical
to the comoving distance, i.e., $r(\chi) = \chi$.
The convergence field is the tracer of the cosmic matter density field,
and thus, the convergence map is also called a ``mass map''.
At the peak positions of high convergence, massive structures such as galaxy clusters
are likely to be located.
Hence, the peaks of the convergence map are used to search massive clusters \citep[e.g.,][]{Miyazaki2018,Oguri2021,Chen2025}.

In the actual observations, the isotropic deformation due to the convergence is degenerate
with the original size of the source galaxy.
Therefore, only the shear field, which induces the anisotropic pattern,
can be estimated by measuring the shapes of galaxies.
In the WL regime, the observed ellipticity of a galaxy
is the sum of the weak lensing shear and the intrinsic shape:
\begin{align}
    \varepsilon_{\mathrm{obs}} \simeq \gamma + \varepsilon_\mathrm{int}
\end{align}
where $\varepsilon_\mathrm{obs}$ is the observed ellipticity
and $\varepsilon_\mathrm{int}$ is the ellipticity of the intrinsic shape.
The ensemble average of the intrinsic ellipticity is zero
because the intrinsic shapes of galaxies are randomly oriented.
Hence, taking an average of galaxy shapes erases the intrinsic shape:
\begin{equation}
\langle \varepsilon_\mathrm{obs} \rangle = \gamma ,
\end{equation}
where $\langle \cdots \rangle$ denotes the ensemble average.
To estimate the shear field in observations, we take averages of close galaxies
which are gravitationally lensed with the same shear field.
Since the number of close galaxies is finite,
the average of the intrinsic shape is not necessarily zero
and the residual dilutes the WL signal.
This residual noise is called as ``shape noise''
and is the primary contaminant in WL measurements.

Next, let us discuss how to measure the convergence field from the shape measurements.
The shear and convergence fields are related in Fourier space:
\begin{align}
    \tilde{\kappa}(\bm{\ell}) = \frac{\ell^2_1 - \ell^2_2 + 2i\ell_1 \ell_2}{\ell_1^2 + \ell_2^2} \tilde{\gamma}(\bm{\ell}),
\end{align}
where $\bm{\ell} = (\ell_1, \ell_2)$ is the wave-vector in the angular space,
and $\tilde{\kappa}$ and $\tilde{\gamma}$ are the convergence and shear in Fourier space.
The convergence field can be obtained by applying the inverse Fourier transform to $\tilde{\gamma}$.
In practice, the convergence field is directly estimated from galaxy shapes
with Kaiser--Squires (KS) inversion \citep{Kaiser1993,Kaiser1995}.
Similarly to the shear field, the shape noise also affects the estimation
of the convergence field.
In the WL regime, the shape noise $N$ appears in the observed convergence field $\kappa_\mathrm{obs}$
in the additive form:
\begin{equation}
\kappa_\mathrm{obs} = \kappa + N .
\end{equation}
The noise field has clear statistical properties \citep{vanWaerbeke2000};
the mean is zero $\langle N (\bm{\theta}) \rangle = 0$ and
the two-point correlation function is
\begin{equation}
\label{eq:shapenoise}
\langle N(\bm{\theta}) N(\bm{\theta}') \rangle =
\delta_\mathrm{D}^2 (\bm{\theta} - \bm{\theta}') \frac{\sigma^2_\varepsilon}{2 n_\mathrm{g}},
\end{equation}
where $\delta_\mathrm{D}^2$ is the two-dimensional Dirac delta function,
$\sigma^2_\varepsilon$ is the variance of intrinsic shapes,
and $n_\mathrm{g}$ is the number density of source galaxies.
Regardless of the underlying probability distribution of the intrinsic shape,
the mean of the shapes asymptotically follows Gaussian distribution according to the central limit theorem.
The intrinsic variance is not tunable, and thus, the high source number density is critical
to accurately estimate the mass map.
However, conducting deep imaging surveys with higher source number density and wide areas
is challenging due to the limited observing time.
In this work, in order to realise accurate mass map reconstruction without demanding a high source number density,
we utilise the machine learning approaches to \textit{denoise} the observed convergence map,
i.e., remove the shape noise from the observed map.
There is a caveat about the statistical property of the shape noise.
In the real WL measurements, it is possible to estimate the shape noise by randomly rotating
source galaxies, which reflects the spatial variations of the number density of source galaxies.
However, the assumed uniform noise does not depend on any specific source galaxy distribution
except the number density, and thus, the results of denoising WL maps are more generic.

\section{Methods}
\label{sec:methods}

\subsection{Generative models for image-to-image translation}
Our objective is to remove the shape noise from the observed convergence field.
In other words, the goal is to find the optimal mapping of
the noisy convergence field to the noiseless convergence field.
To this end, we apply two different machine learning (ML) approaches:
generative adversarial networks (GANs) and diffusion models (DMs).
These approaches are classified as deep generative models, which learn the probability distribution function
of the target observable given the training data sets.
In particular, the denoising process can be regarded as the image-to-image translation problem,
where the input image is noisy observed convergence maps and the output image is noiseless true convergence maps.
The deep generative models are effective at image-to-image translation problems
and demonstrate high performance in practical tasks such as image inpainting, colourisation, and segmentation.
Recently, these methods have also been applied to the noise reduction task
in astrophysics and cosmology \cite[e.g.,][]{Shirasaki2019,Moriwaki2021,Floss2024}.
Here, we overview the technical basics of GANs and DMs.

\subsubsection{Generative adversarial network}
\label{sec:GAN}

First, we review the algorithm of GAN \citep{Goodfellow2014}.
In particular, our model learns the mapping conditioned on the input convergence map,
and thus, we utilise the conditioned version of GAN, i.e.,
conditional GAN\footnote{In this work,
we employ conditional GAN only and refer to conditional GAN as ``GAN'' for simplicity.}.
GANs consist of two competing networks: the generator $G$,
which produces the fake images conditioned on the given image,
and the discriminator $D$, which distinguishes fake and true images.
At the same time, the discriminator compares the input image with the target image
to determine whether the target image is a real image or a fake image generated by the generator.
Through training with data sets, the model parameters $\theta$
are determined by optimising the Jensen--Shannon divergence and regularisation term:
\begin{equation}
\label{eq:loss_GAN}
\theta = \mathrm{arg} \min_G \max_D \left\{ L_\mathrm{GAN}(G, D) + \lambda L_\mathrm{L1} \right\},
\end{equation}
where $G$ is the generator network, $D$ is the discriminator network,
and $\lambda$ is a regularisation parameter that adjusts the weight of
the two loss functions $L_{\mathrm{GAN}}$ and $L_\mathrm{L1}$.
The loss functions are given as
\begin{align}
\label{eq:GAN_loss}
L_\mathrm{GAN} &= \mathbb{E}_{\bm{x},\bm{y}}[\log D(\bm{x},\bm{y})] +
\mathbb{E}_{\bm{x},\bm{z}}[\log (1 - D(\bm{x},G(\bm{x},\bm{z})))], \\
\label{eq:L1_loss}
L_\mathrm{L1} &= \mathbb{E}_{\bm{x},\bm{y},\bm{z}} \sum_\mathrm{map} |\bm{y} - G(\bm{x},\bm{z})|,
\end{align}
where $\mathbb{E}_{\bm{p}}$ denotes the averaging operation over the variable $\bm{p}$,
$\bm{x}$ and $\bm{y}$ are the input and output data vector, respectively,
and $\bm{z}$ is a random noise added to the lowest layer of the generator.
Although it is possible to learn the transformation from the input to the output without noise $\bm{z}$,
Gaussian noise $\bm{z}$ is added to prevent overfitting to the training data distribution and
induce a diversity of output data.
During training, gradient descent is applied alternately to optimise both $D$ and $G$ in each step.
As described in the original implementation of GANs \citep{Goodfellow2014},
during training of $G$, instead of minimizing $\log(1 - D(\bm{x}, G(\bm{x}, \bm{z})))$,
we minimize $-\log D(\bm{x}, G(\bm{x}, \bm{z}))$.
Furthermore, to slow down the learning rate of $D$,
the loss function is divided by two during the optimisation of $D$.
In summary, we alternately minimize the loss functions $L_G$ and $L_D$, which are defined as
\begin{align}
L_G &= \mathbb{E}_{\bm{x},\bm{z}}[-\log D(\bm{x},G(\bm{x},\bm{z}))] + \lambda L_\mathrm{L1}, \\
L_D &= - \frac{1}{2} \left\{ \mathbb{E}_{\bm{x},\bm{y}} \log D(\bm{x},\bm{y}) + \mathbb{E}_{\bm{x},\bm{z}} \log \{1 - D(\bm{x},G(\bm{x},\bm{z}))\} \right\}.
\end{align}
In this work, we use \texttt{pix2pix} \citep{Isola2017} implemented with
\texttt{PyTorch} \citep{PyTorch} \footnote{\url{https://github.com/junyanz/pytorch-CycleGAN-and-pix2pix}},
which employs GAN to take two-dimensional images as data vectors.
In addition to the loss function originally proposed in \citet{Isola2017}
(Eqs.~\ref{eq:GAN_loss} and \ref{eq:L1_loss}),
we also train the model with alternative loss functions:
LSGAN \citep{Mao2016} and WGAN-gp \citep{Gulrajani2017}.
These loss functions are proposed to improve the stability of training for generic GANs.
However, we find that the model trained with the original loss function yields the best results.
LSGAN stabilizes the training but the denoising results are worse than the original loss fuction.
WGAN-gp leads to unstable training by breaking the balance
between generator and discriminator losses, and the generated images are collapsed.
Therefore, we employ the original loss function as fiducial choice.
The details of the architectures of the networks are described in Appendix~\ref{sec:arch_net_GAN}.

While GANs can generate high-quality images quickly,
they suffer from various issues due to the model structure.
In general, the training of GANs is unstable due to the mode collapse and the vanishing gradient problem
compared with other existing generative models such as variational auto-encoders \citep{Kingma2014b}.
Furthermore, the generated images are less diverse because the generator tends to ignore the noise term
and generate images only from the information of the conditioned images as training proceeds.
Due to this problem, \texttt{pix2pix} does not provide the noise term in the input data vectors.
Instead, the stochasticity appears only in dropouts of the generator network.
However, the output images are still less diverse than other generative models even with the treatment,
that is, GANs fail to learn the probability distribution of the target images.

\subsubsection{Diffusion model}
\label{sec:DM}
DM \citep{Sohl-Dickstein2015,Ho2020,Song2021} has been proposed
in order to circumvent the problems of GANs and enhance the quality of generated data.
It is known that training of DM is more stable and can generate more diverse outputs
from a single input data.

The basic idea of DM as a generative model is to learn the process to
remove the noisy data with deep neural networks.
DM consists of two processes, the forward and reverse processes.
The forward process adds a Gaussian noise to the data
and is repeated $T$ times with the initial data $\bm{y}_0$.
At the step $t \, (t = 1, \ldots, T)$, the forward process is given as
\begin{equation}
\bm{y}_t = \sqrt{\alpha_t} \bm{y}_{t-1} + \sqrt{1-\alpha_t} \bm{z}_t,
\end{equation}
where $\bm{y}_t$ is the data at the step $t$, $\alpha_t$ is the hyper-parameter which schedules the noise amplitude and
$\bm{z}_t$ is the noise vector.
The noise follows the following normal distribution;
\begin{equation}
\bm{z}_t \sim \mathcal{N} (\bm{0}, \bm{I}),
\end{equation}
where $\mathcal{N} (\bm{\mu}, \bm{\Sigma})$ is the normal distribution with the mean $\bm{\mu}$ and
the covariance $\bm{\Sigma}$, and $\bm{I}$ is the identity matrix.
The probability distribution function $q(\bm{y}_{t}|\bm{y}_{t-1})$
of the noise-added data $\bm{y}_t$ given the data at the previous step $\bm{y}_{t-1}$
is expressed as
\begin{equation}
q(\bm{y}_{t}|\bm{y}_{t-1}) = \mathcal{N} (\sqrt{\alpha_{t}} \bm{y}_{t-1}, (1-\alpha_t)I) .
\label{eq:diffusion}
\end{equation}
Since the forward process is a Markov process, the probability distribution function
of the data at the step $t$ given the initial data can be analytically expressed
as the product of distribution functions at previous steps:
\begin{equation}
q(\bm{y}_{t}|\bm{y}_0) = \prod_{t'=1}^t q(\bm{y}_{t'}|\bm{y}_{t'-1}) =
\mathcal{N} (\sqrt{\bar{\alpha}_t} \bm{y}_0, \sigma_t^2 \bm{I}),
\end{equation}
where $\bar{\alpha}_t = \alpha_1 \cdots \alpha_t$ and $\sigma_t = \sqrt{1-\bar{\alpha}_t}$.
The noise schedule $\alpha_t$ and the value of $T$ are determined
so that the final data $\bm{y}_T$ can be well approximated as Gaussian noise.

Next, let us consider the reverse process, which inverts the forward process and
corresponds to the operation to recover $\bm{y}_{t-1}$ from $\bm{y}_t$.
Target data can be generated by applying the reverse process iteratively to the initial Gaussian noise.
However, the reverse process cannot be analytically obtained.
The DM approximates the reverse process with the deep neural networks:
\begin{equation}
q_\theta (\bm{y}_{t-1}|\bm{y}_{t}) =
\mathcal{N} \left(\bm{\mu}_\theta (\bm{y}_t, t), \bm{\Sigma}_\theta (\bm{y}_t, t)\right),
\end{equation}
where the mean $\bm{\mu}_\theta$ and the variance $\bm{\Sigma}_\theta$ are expressed
as the deep neural networks with model parameters $\theta$.
Then, let us consider the consistency of the joint probability distribution function
of the series of the data $(\bm{y}_1, \ldots, \bm{y}_T)$.
The joint probability distributions are given with forward process or reverse process:
\begin{align}
q (\bm{y}_1, \ldots, \bm{y}_T) &= q (\bm{y}_0) \prod_{t=1}^T q (\bm{y}_t | \bm{y}_{t-1}) , \\
q_\theta (\bm{y}_1, \ldots, \bm{y}_T) &= p (\bm{y}_T) \prod_{t=1}^T q_\theta (\bm{y}_{t-1} | \bm{y}_t) ,
\end{align}
where $p (\bm{y}_T)$ is the prior of the final data and taken as unit normal distribution.
The objective is to minimise the Kullback--Leibler divergence of the two distributions:
$q (\bm{y}_1, \ldots, \bm{y}_T)$ and $q_\theta (\bm{y}_1, \ldots, \bm{y}_T)$.

In practice, for better convergence of the training process and sampling efficiency,
\citet{Ho2020} proposed the simplified scheme;
the objective function for optimisation is the reweighted variational lower-bound
and the covariance is diagonal $\bm{\Sigma}_\theta = \sigma_t^2 \bm{I}$.
Finally, the loss function $L_t$ for the step $t$ is given as
\begin{equation}
\label{eq:loss_DM}
L_t = \mathbb{E}_{\bm{x},\bm{y}} \mathbb{E}_{\bm{\epsilon} \sim \mathcal{N} (\bm{0},\bm{I})}
\left| \bm{\epsilon}_\theta \left( \sqrt{\bar{\alpha}_t} \bm{y}_t + \sqrt{1-\bar{\alpha}_t} \bm{\epsilon} , t \right)
- \bm{\epsilon} \right|^2,
\end{equation}
where the noise vector $\bm{\epsilon}_\theta$ is related with $\bm{\mu}_\theta$ as
\begin{equation}
\bm{\mu}_\theta (\bm{y}_t, t) = \frac{1}{\sqrt{\alpha_{t}}}
\left(\bm{y}_{t}-\frac{1-\alpha_{t}}{\sqrt{1-\bar{\alpha}_{t}}}
\bm{\epsilon}_{\theta}(\bm{y}_{t},t)\right) .
\end{equation}
Then, the loss functions are optimised by uniformly but randomly
iterating over steps $t = 1, \cdots, T$.

The generation process begins with a Gaussian noise $\bm{y}_T$
and samples the denoised data with the learned reverse process:
\begin{align}
\bm{y}_{t-1} = \frac{1}{\sqrt{\alpha_t}}\left( \bm{y}_t - \frac{1-\alpha_t}{\sqrt{1-\bar{\alpha_t}}} \bm{\epsilon}_\theta (\bm{x},\bm{y}_t,t) \right) + \sqrt{1-\alpha_t}\bm{\epsilon},
\end{align}
where $\bm{\epsilon} \sim \mathcal{N} (0, \bm{I})$.
Iterating the reverse process $T$ times, we obtain the generated data in the target space.
The stochasticity is introduced in the initial data $\bm{y}_T$ and
the noise at each step $\bm{\epsilon}$.
These elements are the key to generating diverse data in DM.

In this work, we employ \texttt{Palette} \citep{Saharia2021} model, which implements
DM to take images as inputs and outputs with \texttt{PyTorch} \citep{PyTorch}.
Similarly to GANs, \texttt{Palette} takes the input image $\bm{x}$ used as the conditioned image,
and thus, the model should be referred to as ``conditional'' DM.
Conditional DM can be implemented by adding the input image
as the argument of the noise vector, i.e.,
\begin{equation}
\bm{\epsilon}_\theta \left( \sqrt{\bar{\alpha}_t} \bm{y}_t + \sqrt{1-\bar{\alpha}_t} \bm{\epsilon} , t \right)
\to
\bm{\epsilon}_\theta \left( \sqrt{\bar{\alpha}_t} \bm{y}_t + \sqrt{1-\bar{\alpha}_t} \bm{\epsilon} , \bm{x}, t \right) .
\end{equation}
The details of the architectures of the networks are described in Appendix~\ref{sec:arch_net_DM}.

DM can generate more high-quality and diverse data compared with GAN.
Furthermore, the training process of DM is more stable
because of the simple objective function (Eq.~\ref{eq:loss_DM})
in contrast to competing loss functions of the generator and discriminator in GAN.
The drawback of DM is the computational cost of data generation.
In general, the number of total diffusion steps $T$ is required to be $T \gtrsim \mathcal{O}(10^3)$,
which takes considerably longer than GANs.

\subsection{Simulation data and survey realisms}
\label{sec:simulation_data}
In order to evaluate the performance of denoising with GAN and DM,
we generate realistic mock convergence maps as pairs of maps with and without shape noise.
To this end, we employ the $\kappa$TNG mock WL data suite \citep{Osato2021},
which consists of 10,000 pseudo-independent mock weak lensing maps
with the wide source redshift range ($0.03 \leq z_s \leq 2.6$)
based on the multiple plane ray-tracing approach \citep{Hilbert2009}.
The large number of realisations is suitable for training data for ML-based approaches.
In this work, we employ the convergence maps with the source redshift $z_s = 1$
unless otherwise stated.
The convergence map of $\kappa$TNG is square shaped with the area of $5 \times 5\, \mathrm{deg}^2$
and pixellated with regular $1024$ grids per dimension.
Though the main suite of $\kappa$TNG is generated from IllustrisTNG
hydrodynamical simulations with galaxy formation physics,
we use the other suite, $\kappa$TNG-Dark,
simulated with IllustrisTNG-Dark dark-matter only simulations
for a fair comparison with previous studies using dark-matter only simulations.
Hereafter, we refer this $\kappa$TNG-Dark suite to $\kappa$TNG for simplicity.

For efficiency and stability of numerical computations, we post-process $\kappa$TNG convergence maps.
First, to increase the number of maps, we divide each map into 4 parts with equal area.
The size of the map after division is halved, i.e., $2.5 \times 2.5\, \mathrm{deg}^2$
with $512$ grids per dimension.
Next, to ease the computational costs,
we reduce the number of grids from $512$ to $256$
by average pooling with the pooling size $2 \times 2$.
The final data set consists of 40,000 maps with $2.5 \times 2.5\, \mathrm{deg}^2$
with $256$ grids per dimension. The angular grid size is $0.59 \, \mathrm{arcmin}$.
Since we smooth the maps with a large smoothing scale,
the pixel size is small enough to study the denoising process.

Since the original mock convergence maps are noiseless, we add the shape noise manually.
The amplitude of the shape noise is determined in Eq.~\eqref{eq:shapenoise},
and we adopt the intrinsic variance $\sigma_\varepsilon = 0.35$ and
the source number density $n_\mathrm{gal} = 20 \, \mathrm{arcmin}^{-2}$,
which are similar to the Subaru HSC weak lensing
survey \citep{Aihara2018}\footnote{We present denoising results with higher source number density in Appendix~\ref{sec:COSMOS_mock}}.
Then, we assign the noise value from the Gaussian distribution with a mean of zero
and a variance of $\sigma_\mathrm{pix}^2$ to each pixel independently,
and the variance per pixel is given as
\begin{equation}
\label{eq:sigma_pix}
\sigma_\mathrm{pix}^2 = \frac{\sigma_\varepsilon^2}{2 A_\mathrm{pix} n_\mathrm{g}},
\end{equation}
where $A_\mathrm{pix}$ is the angular area of the pixel of the maps.
Since it is expected the shape noise and weak lensing effect are independent,
we simply add the shape noise to the true convergence map.
The mock observed convergence map $\kappa_\mathrm{obs}$ is given as
\begin{align}
    \kappa_{\mathrm{obs}} = \kappa_\mathrm{true} + N
\end{align}
where $\kappa_\mathrm{true}$ is the true $\kappa$TNG convergence map without shape noise,
and $N$ is the shape noise sampled from the normal distribution with zero mean and
the variance of $\sigma_\mathrm{pix}^2$.
Note that the noise realisation is independent among 40,000 mass maps.
In the simulated maps, survey masks are not considered.
However, in the real analysis, some regions are masked due to bright stars or instrumental defects.
In addition, observational systematics may distort the probability distribution function of the shape noise
from the Gaussian distribution. Such effects are beyond the scope of this work, and we focus on the comparison between GAN and DM in the ideal case.

In practice, to reduce the noise at small scales and transients due to instrumental systematics,
the isotropic two-dimensional Gaussian filter is applied to the maps:
\begin{align}
    \kappa_\mathrm{sm} (\bm{\theta}) &= \int d^2 \theta' \, \kappa(\bm{\theta'}) W_\mathrm{G} (\bm{\theta'} - \bm{\theta}),\\
    W_\mathrm{G} (\bm{\theta}) &= \frac{1}{\pi \theta_\mathrm{G}^2} \exp\left(-\frac{|\bm{\theta}|^2}{\theta_\mathrm{G}^2}\right),
\end{align}
where $\kappa_\mathrm{sm}$ is the smoothed convergence map,
$W_\mathrm{G} (\bm{\theta})$ is the Gaussian filter and $\theta_\mathrm{G}$ is the smoothing scale.
Hereafter, we adopt $\theta_\mathrm{G} = 1.5\, \mathrm{arcmin}$,
which is an optimal smoothing scale for the detection of
massive galaxy clusters at $z_s \simeq 1$ \citep{Miyazaki2018}.
All the relevant fields are smoothed with this filter unless otherwise stated.
In summary, our goal is to reconstruct the underlying convergence field
$\kappa_\mathrm{true}$ from the observed noisy convergence field $\kappa_\mathrm{obs}$.

\subsection{Training and testing processes}
\label{sec:training}
Here, we describe the training and testing processes of GAN and DM.
First, we generate 40,000 independent noise map sampled from a Gaussian distribution
and add them to $\kappa$TNG maps.
Then, we randomly split the 40,000 $\kappa$TNG maps into 39,000 maps for training and 1,000 maps for testing.
Finally, 40,000 pairs of noiseless and noisy convergence maps are created.

In the previous work with GAN \citep{Shirasaki2019}, 40,000 noisy convergence maps are created with
bootstrapping from 200 noiseless convergence maps while noise realisations are independent
among 40,000 maps.
Following \citet{Shirasaki2019}, we generate 40,000 maps bootstrapped from 200 realisations
randomly selected from the original 40,000 $\kappa$TNG maps.
However, it turns out that the denoising performance is better
when the maps without bootstrapping are used, i.e., all 40,000 convergence maps are independent.
Therefore, in this work, we use 40,000 independent convergence maps as fiducial data set.

In order to reduce the dynamic range of the data and suppress numerical errors,
we apply the scaling to the noisy observed map
$\kappa_\mathrm{obs}$ and the true noiseless convergence map $\kappa_\mathrm{true}$ for DM
to match with the range of the noise amplitude (Eq.~\ref{eq:diffusion}).
The noisy and noiselss maps are rescaled as
\begin{equation}
\label{eq:preprocess}
    \kappa \to \frac{\kappa}{\kappa_\mathrm{max}},
\end{equation}
where $\kappa_\mathrm{max}$ is the maximum value of the entire data set
and is estimated for $\kappa_\mathrm{obs}$ and $\kappa_\mathrm{true}$ separately.
Note that using the data set without rescaling can also generate
the denoised convergence maps at some level,
but this amplitude rescaling improves the denoising performance significantly.

For training of GAN, we adopt the regularisation parameter $\lambda = 100$
in Eq.~\eqref{eq:loss_GAN} with a batch size of 1, which is shown to attain better performance
for the generator with a U-Net architecture \citep[see, e.g., Appendix~6.2 in][]{Isola2017}.
We use Adam optimiser \citep{Kingma2014a} for training.
The initial learning rate is set to 0.0002, and after 100 epochs,
the learning rate decays linearly to 0 for additional 100 epochs.
In total, the model is trained for 200 epochs.
We train the model with a single NVIDIA A100 GPU, and it took approximately 28 hours.
In contrast, testing phase, i.e., generating 1,000 denoised maps,
finishes within a few minutes.

For DM, the model has several hyper-parameters,
e.g., diffusion steps, sampling steps, noise schedule, and noise variance,
which are specific to its model framework.
These hyper-parameters are crucial for generating high-quality outputs \citep{Nichol2021,Chen2023}.
We evaluate the performance with one of hyper-parameters changed from the fiducial one
and find that the diffusion steps affect the accuracy of the denoising process the most.
Based on these tests, the following hyper-parameters are adopted.
The model is trained for 85 epochs with a batch size of 4 and Adam optimiser.
The learning rate is set to $5 \times 10^{-5}$.
The timestep of the diffusion process is set to 4,000,
and we adopt a quadratic noise schedule as the noise scheduling (Eq.~\ref{eq:diffusion}).
Among hyper-parameters, the noise scheduling significantly affects the results at small scales.
The in-depth comparison of the noise scheduling is presented in Appendix~\ref{sec:noise_schedule}.
We use the same NVIDIA A100 GPU,
and it takes approximately 45 hours for training.
For the testing process, the variance is divided in the same manner as in training,
with a timestep of 2,000 and a batch size of 64.
Unlike GANs, the testing for 1,000 maps takes approximately 6 hours,
i.e., 22 seconds per map with the same GPU.
It should be noted that increasing the number of timesteps results in a longer testing time.
For instance, setting the timestep to 4,000 requires approximately 12 hours.

In order to virtually introduce diversity in GAN,
we train GAN model parameters five times with different initialisations.
As a result, the five generator networks denoise the noisy maps in a different manner,
and there are five denoised maps from a single noisy map.
In contrast, DM can sample five times by sampling the initial noise vector with a network trained once.
In summary, from the 1,000 test maps, we generate 5,000 denoised maps with five generator networks,
while DM produces 5,000 denoised maps by sampling five times using the single model.
There is one caveat about the variance among 5 samples in GAN and DM.
The meaning of the variance among 5 samples is fundamentally different.
The variance among the five samples of GANs should be regarded as the uncertainty;
the determined network weights are not necessarily optimal but fluctuate.
In contrast, DM learns the probability distribution function of denoised maps given the noisy map,
and thus, the variance of 5 DM samples reflects the learned probability distribution function.
We also conduct retraining DM with different initial random seed as done in GAN
but the difference in terms of statistics is negligible,
which also demonstrates the stability of the training process of DM.

For denoising the WL mass maps, it is naively expected that
the input map is the noisy observed map $\kappa_\mathrm{obs}$
and the output map is the noiseless true map $\kappa_\mathrm{true}$,
which we refer to as direct transformation.
However, the previous study \citep{Shirasaki2019} found better performance with the indirect translation,
where the output map is the noise map $N$, and the denoised map is obtained
by subtracting the estimated noise map from the observed noisy map.
We show the comparison of direct and indirect transformation in Appendix~\ref{sec:indirect}
and find that the direct transformation yields better results in our setting.
Therefore, the direct transformation is adopted as the default in this paper.

\section{Results}
\label{sec:results}

\subsection{Visual inspection}
\begin{figure*}
  \begin{center}
  \includegraphics[width=17cm]{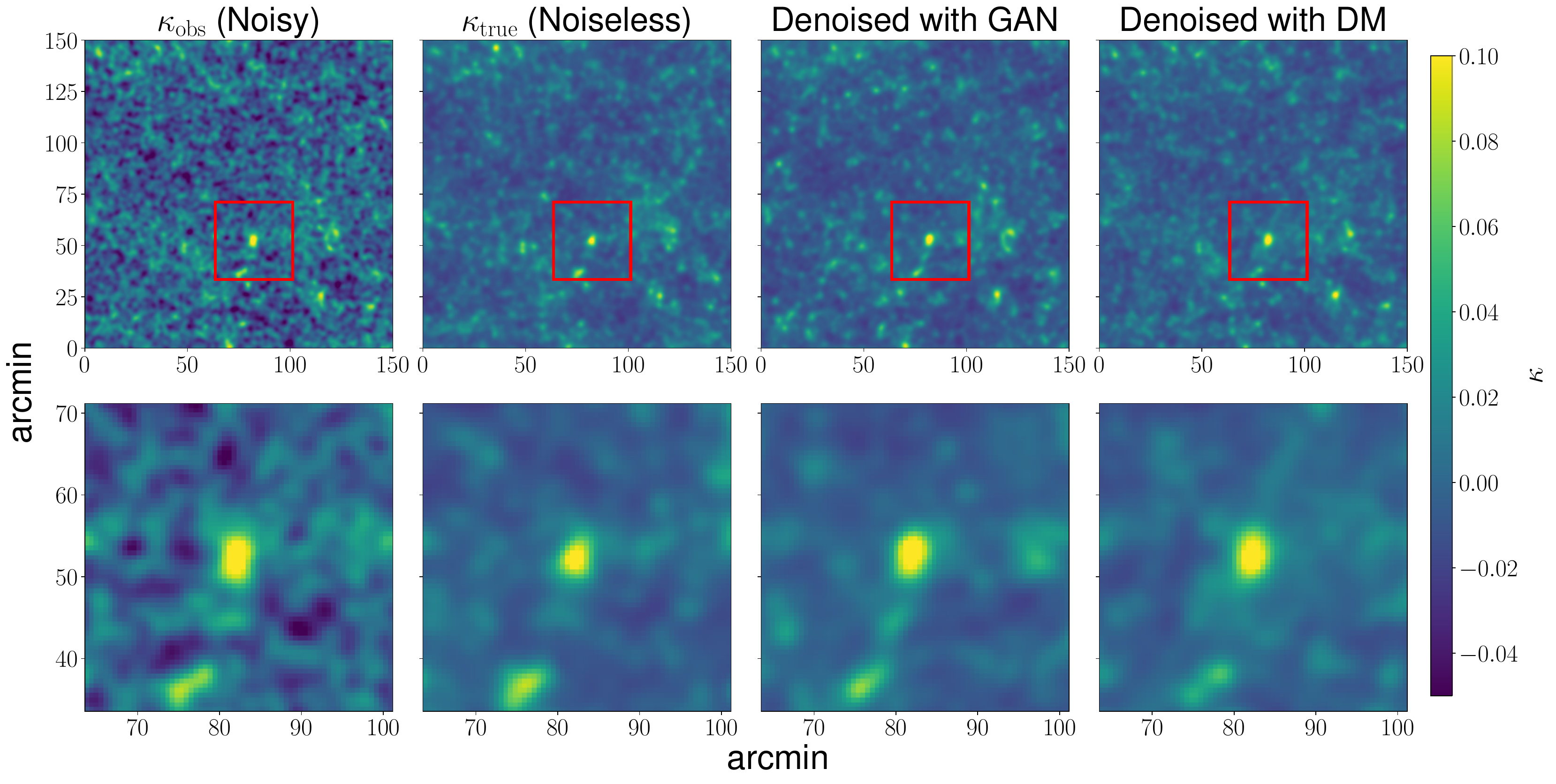}
  \end{center}
  \caption{\textit{Upper row}: Comparison between the denoised convergence maps
  with GAN and DM and the ground truth map.
  These maps are pixellated with $256 \times 256$ grids with the area of $2.5 \times 2.5 \deg^2$,
  which corresponds to the pixel size of $0.586\,\mathrm{arcmin}$.
  \textit{Lower row}: Zoom-in maps around the peak with the highest significance in the true noiseless map.
  The red squares in the maps in the upper row indicate
  the corresponding zoom-in regions displayed in the lower row.
  Note that each map is not normalised.
  {Alt text: Eight colour images.}}
\label{fig:maps}
\end{figure*}

First, we visually examine the denoised convergence maps generated by GAN and DM.
Figure~\ref{fig:maps} shows one realisation of
the noisy convergence map $\kappa_\mathrm{obs}$,
the true noiseless convergence map $\kappa_\mathrm{true}$,
the denoised maps with GAN and DM from the test set of 1,000 realisations.
Note that normalisation is not applied to the maps in Figure~\ref{fig:maps}.
Overall, both GAN and DM can remove noisy features at small scales and reconstruct the mass map at large scales.
The bottom panels of Figure~\ref{fig:maps} display zoom-in views
around the highest peak in the true noiseless convergence map.
The morphology of the peak in denoised maps with GAN and DM is different,
but the peak remains after denoising.
However, not all peaks in noiseless maps, in particular peaks with lower heights,
cannot be reconstructed in denoised maps with GAN and DM, and even spurious peaks,
which do not have corresponding peaks in the noiseless map, appear in the denoised maps.
This trend in peak identification has also been reported
in previous studies using GAN \citep{Shirasaki2019}.
The in-depth discussions on the effects of denoising on peaks are presented in Section~\ref{sec:peak_recon}.

\begin{figure*}
  \begin{center}
  \includegraphics[width=17cm]{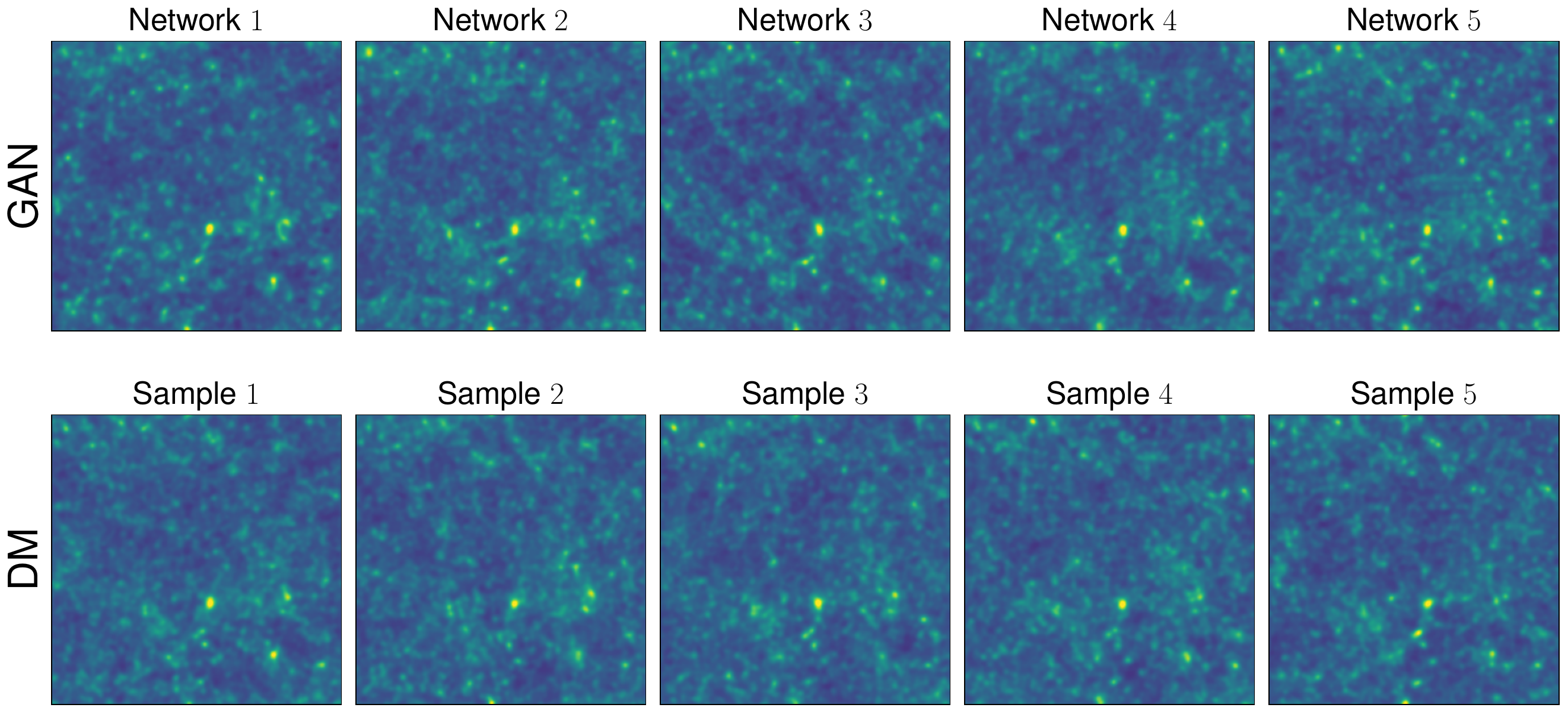}
  \end{center}
  \caption{Comparison of the denoised maps generated by five networks of GAN and five samples of DM
  from a single noisy map.
  {Alt text: Ten colour images.}}
  \label{fig:maps_sample}
\end{figure*}

\begin{figure}
  \begin{center}
  \includegraphics[width=\columnwidth]{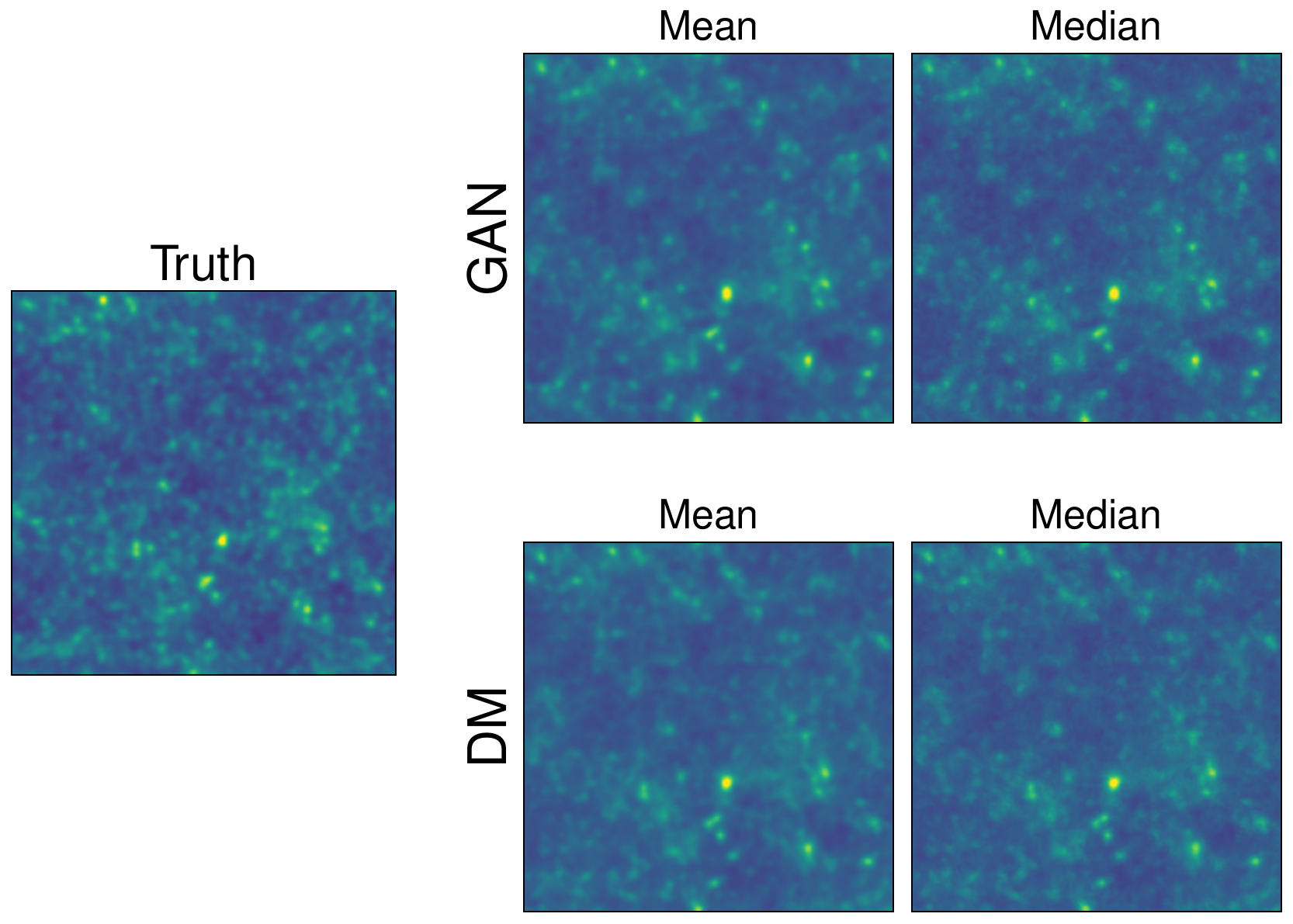}
  \end{center}
  \caption{Comparison between the true noiseless map and the mean and median
  of five denoising realisations with GAN and DM in Figure~\ref{fig:maps_sample}.
  {Alt text: Five colour images.}}
  \label{fig:map_mean}
\end{figure}

As described in Section~\ref{sec:training},
GAN and DM generate five denoised maps from a single observed map.
Figure~\ref{fig:maps_sample} presents five samples with GAN and DM
for the same realisation of the convergence field in Figure~\ref{fig:maps}. 
Though a slight difference can be seen especially around peaks,
there is no significant difference in the overall structures among the
five samples.
In addition, Figure~\ref{fig:map_mean} shows the mean and median convergence maps obtained from the five denoised maps
in Figure~\ref{fig:maps_sample}.
These maps indicate smooth features, which suggests that taking a mean or median smears out the small-scale transients.
In particular, in this specific realisation,
the standard deviation of the averaged map is different by $5\%$ from the ground truth in the individual samples,
but this deviation increases to about $10\text{--}20\%$ in the averaged maps.

\subsection{Peak location reconstruction}
\label{sec:peak_recon}
Next, we investigate the difference between the locations of the peaks before and after denoising.
To characterise the peak height, we define the signal-to-noise ratio (SNR) as
\begin{equation}
\SNR \equiv \frac{\kappa - \mu}{\sigma},
\end{equation}
where $\mu$ and $\sigma$ are the mean and standard deviation at the pixel level, respectively.
The lower panel of Figure~\ref{fig:maps} illustrates the convergence map around the very high peak ($\SNR \sim 10$),
and the location of such high peaks does not change after denoising.
However, a large fraction of lower peaks are sourced from the noise transients, and thus, such peaks can be
erased by denoising models even though the peaks are caused by less massive dark matter halos.
Therefore, we focus on peaks with heights $\SNR > 4$, which are likely to have the physical origin,
in this analysis.
The regions close to the edge are subject to incomplete smoothing,
and thus, we exclude the peaks near the edges within $2\theta_\mathrm{G}$.

First, we discuss the origins of the peaks in the \textit{denoised} maps.
In this analysis, for each test map, we use one of five networks for GAN and one of five samples for DM.
For each peak in the denoised map, we search for corresponding peaks
which are located within $2\theta_\mathrm{G} = 3\, \mathrm{arcmin}$ with $\SNR > 4$ in the corresponding noiseless true map.
If there is at least one peak which satisfies the criterion, the peak is classified
as the peak with counterparts.
Table~\ref{tab:peak_denoised} shows the number of total peaks in the denoised map,
peaks with counterparts in the corresponding noiseless map
and fake peaks, i.e., peaks with no counterparts in the true map.
The total numbers of peaks in denoised maps with GAN and DM are quite similar at the $10\%$ level,
and roughly $55\%$ of peaks have counterparts in the true maps.
The rate of peaks with counterparts is slightly higher for DM than GAN.

\begin{table}
    \tbl{The average numbers of total peaks, peaks with counterparts, and fake peaks per a denoised map.
    The numbers are measured in all 5,000 denoised maps with GAN and DM.}
    { 
    \begin{tabular}{l|cc}
        \hline
        & GAN & DM \\
        \hline
        \hline
        Peaks in denoised maps & 20.22 & 17.65 \\
        Peaks with counterparts & 10.95 (54.9\%) & 10.02 (55.9\%) \\
        Fake peaks & 9.26 (45.1\%) & 7.77 (44.1\%) \\
        \hline
    \end{tabular}
    }
    \label{tab:peak_denoised}
\end{table}

Next, we focus on the peaks in the \textit{true} mass maps.
In this case, we divide peaks into two classes: ``surviving peaks'' which have the counterparts even after denoising
and ``missing peaks'' which lose the counterparts in the denoised map.
The definition of counterparts is the same as the previous case: the peaks within $2 \theta_\mathrm{G}$
with $\SNR > 4$ in the corresponding denoised map.
Table~\ref{tab:peak_true} shows the numbers of surviving and missing peaks.
Similarly to the previous case, the difference between GAN and DM is minor.
For both models, $60\%$ of peaks in the true maps remain in the denoised maps
and $40\%$ of peaks are missing in the denoised peaks for several reasons:
the peak height goes below the threshold ($\SNR = 4$) or the peak position is misplaced.
The overall performance is slightly better for GAN than DM.

\begin{table}
    \tbl{The average numbers of total peaks, peaks with counterparts, and fake peaks per a true map.
    The numbers are measured in all 1,000 true maps.}
    {
    \begin{tabular}{l|cc}
        \hline
        & GAN & DM \\
        \hline
        \hline
        Peaks in true maps & \multicolumn{2}{|c}{17.86} \\
        Surviving peaks & 10.94 (61.8\%) & 9.78 (55.2\%) \\
        Missing peaks & 6.92 (38.2\%) & 8.08 (44.8\%) \\
        \hline
    \end{tabular}
    }
    \label{tab:peak_true}
\end{table}

Finally, in order to quantify the gain of denoising, we measure the performance of
the peak location reconstruction without denoising.
Only in this case, the denominator of the peak SNR is the standard deviation of the corresponding noiseless map
instead of that of the noisy map, because the variance of the noisy map is amplified by the shape noise.
Thus, we normalise noisy maps with the standard deviation of the noiseless convergence map.
Table~\ref{tab:peak_noisy} shows the reconstruction results for the noisy maps.
The shape noise leads to more fake peaks, and the total number of peaks is doubled compared with
the case with denoising. $39.2\%$ of peaks have counterparts in the corresponding true map.
Since $54.9\%$ and $55.9\%$ have counterparts for maps denoised with GAN and DM, respectively,
and fewer fake peaks are generated, denoising has a practical gain in peak reconstruction.

\begin{table}
    \tbl{The average numbers of total peaks, peaks with counterparts, and fake peaks per a noisy map.
    The numbers are measured in all 1,000 noisy maps.}
    {
    \begin{tabular}{l|cc}
        \hline
        & Noisy \\
        \hline
        \hline
        Peaks in noisy maps & 43.05  \\
        Peaks with counterparts & 16.19 (39.2\%) \\
        Fake peaks & 26.86 (60.8\%) \\
        \hline
    \end{tabular}
    }
    \label{tab:peak_noisy}
\end{table}

\subsection{Pixel-level comparison}
\begin{figure*}
  \begin{center}
  \includegraphics[width=17cm]{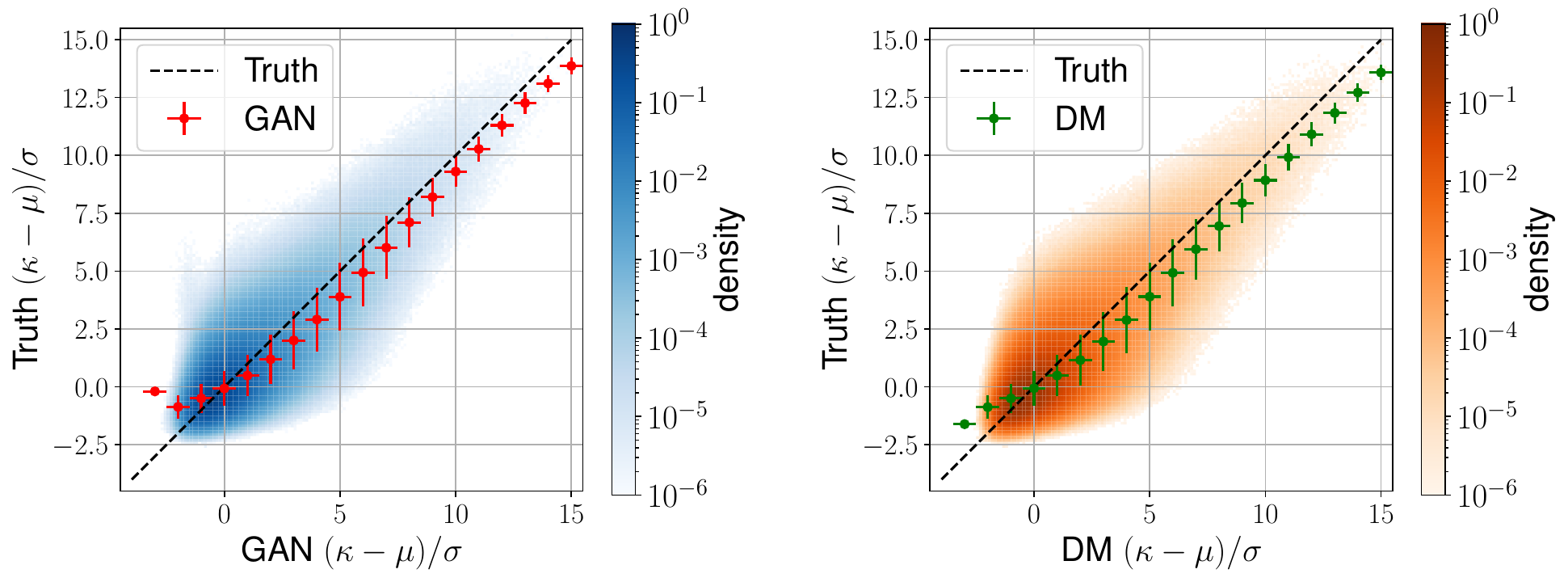}
  \end{center}
  \caption{Pixel-level comparisons of SNR
  between the ground truth map and the denoised maps with GAN (left panel) and DM (right panel).
  The red (green) points with error bars correspond to the mean and standard deviation binned with pixel SNRs in
  5000 denoised maps with GAN (DM).
  The black dashed line represents the case where the generated values perfectly match the ground-truth.
  {Alt text: Two scatter plots.}}
  \label{fig:pixel}
\end{figure*}

Here, we compare the denoised convergence maps with GAN and DM
with the corresponding ground-truth convergence maps on a pixel basis.
Figure~\ref{fig:pixel} represents the relationship between the denoised and corresponding true maps.
Following \cite{Shirasaki2019}, the convergence maps are normalised so that the mean is zero and the variance is unity.
This normalisation allows us to discern small differences between the denoised and true maps clearly.
We exclude the regions near the edges within $2\theta_\mathrm{G}$ due to incomplete smoothing.
The noisy and denoised maps are strongly correlated, and the central value is close to the correct pixel value,
though there is a statistical fluctuation.
Both GAN and DM reproduce the true map within $1\sigma$ for the wide range $-1 \lesssim \SNR \lesssim 9$.

We also investigate the quantitative pixel-level difference between the denoised and ground-truth maps
and measure the root mean square error (RMSE) and the Pearson correlation coefficient $\rho$.
RMSE is defined as
\begin{equation}
\label{eq:RMSE}
    \mathrm{RMSE} \equiv \sqrt{\frac{1}{N^2} \sum^N_{i, j = 1}
    \left(
    \kappa^{\mathrm{unnorm}}_{\mathrm{denoised}} (\bm{\theta}_{ij}) -
    \kappa^{\mathrm{unnorm}}_\mathrm{true} (\bm{\theta}_{ij})\right)^2},
\end{equation}
where $\bm{\theta}_{ij}$ is the angular position of the pixel,
the subscript $i$ ($j$) denotes the pixel index along $x$-axis ($y$-axis),
and $N = 244$ is the number of grids per dimension after excluding the region near the edges.
Note that only in RMSE, unnormalised raw convergence maps are used.
Next, Pearson correlation coefficient $\rho$ is computed as
\begin{equation}
\label{eq:Pearson}
    \rho = \frac{1}{N^2}\sum^N_{i, j = 1}
    {\kappa_\mathrm{denoised} (\bm{\theta}_{ij})}
    {\kappa_\mathrm{true} (\bm{\theta}_{ij})}.
\end{equation}
Table~\ref{tab:RMSE_Pearson} shows the RMSE and Pearson coefficients
of 1,000 mean or median maps of five denoising realisations
and 5,000 individual maps estimated by GAN and DM.
First, we measure the RMSE and Pearson coefficients for the noisy maps without denoising,
which are labelled as ``No denoising'' in Table~\ref{tab:RMSE_Pearson}.
RMSE significantly improves by denoising with GAN or DM,
which demonstrates the effectiveness of the denoising methods.
On the other hand, the Pearson coefficient computed from individual maps stays similar even after denoising.
Taking mean or median of the maps among the five realisations results in improvement for both metrics.
In terms of comparison between GAN and DM,
both methods show similar results in RMSE and Pearson correlation coefficient
within $1\%$ level.

\begin{table}
    \tbl{RMSE and Pearson correlation coefficients $\rho$ for the denoised maps with GAN and DM.
    These quantities are measured for five individual samples
    and their mean and median.
    The metrics computed from noisy maps before applying denoising are also shown as ``No denoising''.
    As indicated by the arrows, the smaller RMSE and larger Pearson correlation coefficients correspond to a better match with the true map.
    The boldface denotes the best value among the individual, mean, and median maps.}
    {
    \begin{tabular}{l|cc|cc}
        \hline
        & \multicolumn{2}{c|}{RMSE ($\times 10^{-2}$) $\downarrow$} & \multicolumn{2}{c}{Pearson coef. $\rho$ $\uparrow$} \\
        & GAN & DM & GAN & DM \\
        \hline
        \hline
        Individual & 1.12 & 1.11 & 0.644 & 0.638\\
        Mean & 0.87 & \textbf{0.86} & \textbf{0.758} & 0.757 \\
        Median & 0.90 & 0.89 & 0.743 & 0.742 \\
        \hline
        \hline
        No denoising & \multicolumn{2}{c|}{1.47} & \multicolumn{2}{c}{0.67}\\
        \hline
    \end{tabular}
    }
    \label{tab:RMSE_Pearson}
\end{table}

\subsection{Statistics}

\subsubsection{Power spectrum and cross-correlation}
Next, we investigate the cosmological statistics to quantify the performance of denoising.
We begin with the angular power spectrum $C (\ell)$, which is defined as
\begin{equation}
\langle \tilde{\kappa}(\bm{\ell}) \tilde{\kappa}(\bm{\ell}')\rangle \equiv (2\pi)^2\delta_\mathrm{D} (\bm{\ell} + \bm{\ell}')C(\ell),
\end{equation}
where $\tilde{\kappa}$ denotes the convergence field in Fourier space,
$\langle \cdots \rangle$ is the ensemble average,
and $\bm{\ell}$ is the multipole.
Since the convergence maps are normalised with the variance,
we define the normalized power spectrum \(\tilde{C}(\ell)\) as
$\tilde{C}(\ell) = C(\ell)/\sigma^2$,
where $\sigma^2$ is the variance of the convergence map at the pixel level.
The power spectra are computed with 20 bins equally spaced in logarithmic space
for the range $\ell \in [200, 10^4]$.

\begin{figure}
  \begin{center}
  \includegraphics[width=\columnwidth]{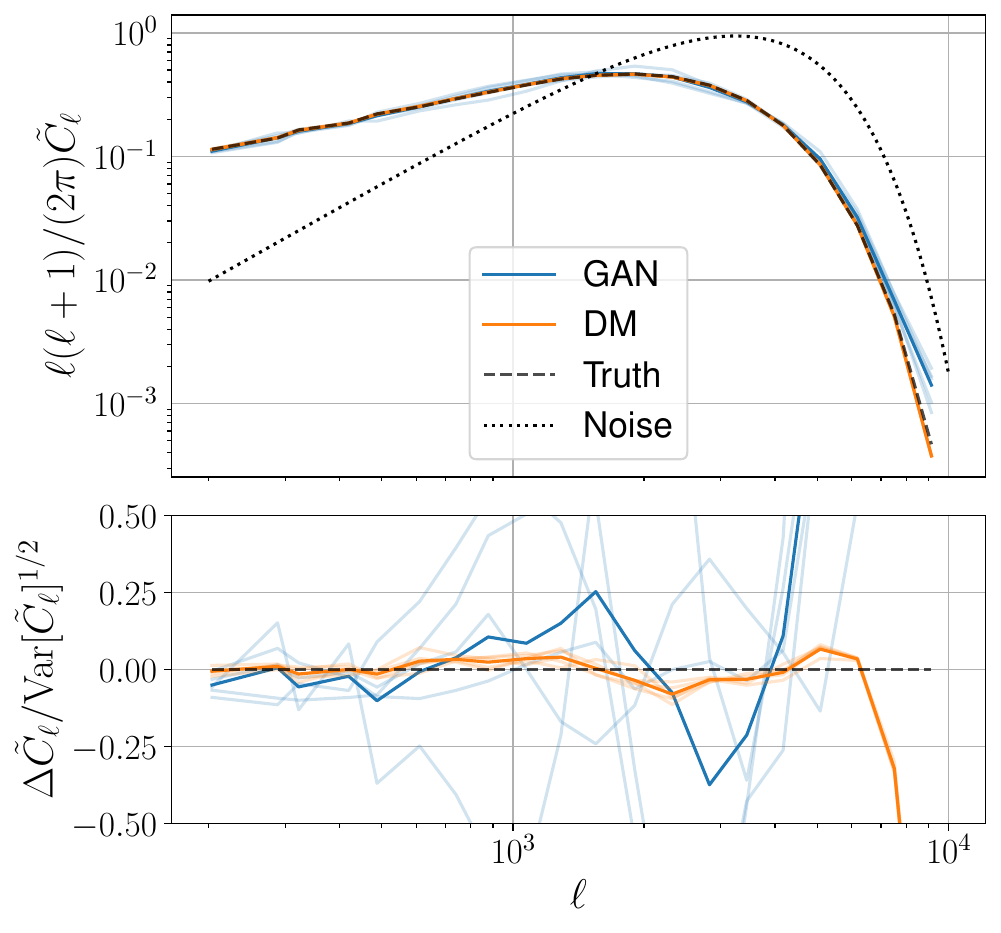}
  \end{center}
  \caption{The angular power spectra of the normalised ground-truth maps (dashed line) and denoised maps (solid lines) with GAN and DM.
  The five blue (orange) thin lines correspond to the five networks (samples) from GAN (DM).
  The thick blue (orange) line shows the average power spectrum for GAN (DM).
  The dotted line corresponds to the noise power spectrum.
  For normalisation of the noise power spectrum, the harmonic average of the variance of true maps is used.
  The bottom panel shows the error between the power spectra of the denoised and the true maps.
  {Alt text: Four lines graph in the upper panel and three lines graph in the lower panel.}}
  \label{fig:cl}
\end{figure}

Figure~\ref{fig:cl} shows the power spectra of noiseless convergence maps and denoised maps with GAN and DM,
and the power spectrum of each sample is shown as thin lines.
The lower panel illustrates the fractional normalised difference, which is defined as
\begin{equation}
\Delta \tilde{C}(\ell) / \mathrm{Var}[\tilde{C}(\ell)]^{1/2} \equiv
\frac{\tilde{C}_\mathrm{denoised}(\ell) - \tilde{C}_\mathrm{truth}(\ell)}
{\mathrm{Var}[\tilde{C}_\mathrm{truth}(\ell)]^{1/2}},
\end{equation}
where $\mathrm{Var}[\tilde{C}_\mathrm{truth}(\ell)]$ is the variance of
the normalised true power spectra over 1,000 maps.
As described in Section~\ref{sec:training}, there are five samples from GAN and DM from the single noisy map,
and the thin solid lines correspond to the mean power spectra of 1,000 generated maps for each of the five samples.
The thick solid lines correspond to the average of all 5,000 maps.
Overall, DM outperforms GAN with regard to power spectrum reconstruction;
DM can reproduce the power spectrum within $0.1\sigma$ for $\ell \lesssim 6000$.
Though the noise power spectrum dominates for $\ell \gtrsim 2000$,
DM accurately reconstructs the power spectrum at such scales.
On the other hand, the range where GAN reproduces the correct power spectrum
is limited to the large scale ($\ell \lesssim 1000$).
Furthermore, the variance among the five networks for GAN is quite large even at the intermediate scale ($\ell \sim 1000$).
Averaging over five samples results in a reasonable estimate of the power spectrum
but the deviation for some models reaches $0.5 \sigma$.
In contrast, the five samples of DM indicate less variance;
all five samples perform well up to $\ell \simeq 6000$.

\begin{figure}
  \begin{center}
  \includegraphics[width=\columnwidth]{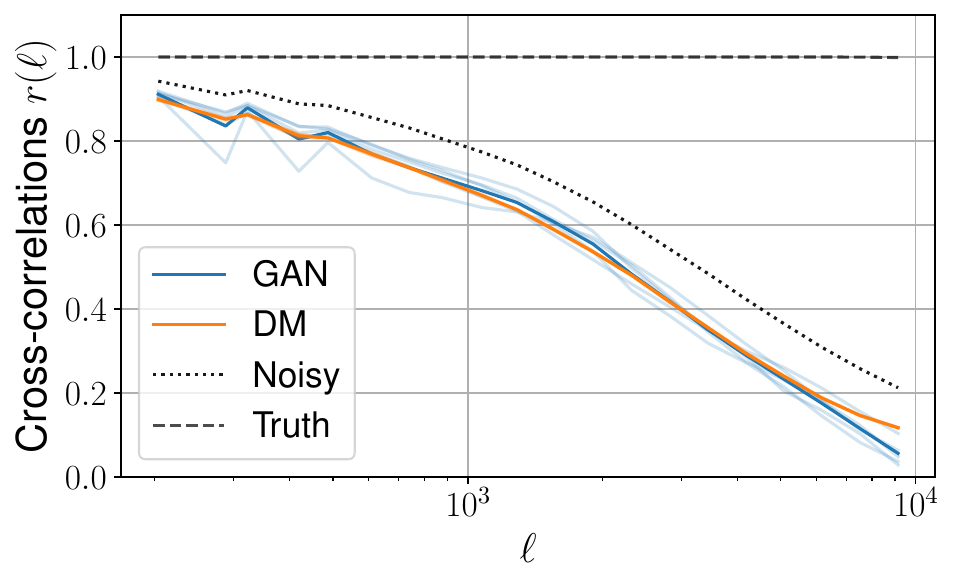}
  \end{center}
  \caption{Cross-correlation coefficients in Fourier space of the convergence maps
  denosied with GAN and DM and the ground-truth.
  The five blue (orange) thin lines correspond to the five networks (samples) from GAN (DM).
  The dotted line corresponds to the case of the noisy maps.
  The black dashed line indicates the case of perfect correlation.
  {Alt text: Two line graph.}}
  \label{fig:cross_cl}
\end{figure}

Figure~\ref{fig:cross_cl} shows the cross-correlation coefficient in Fourier space $r(\ell)$
between the ground-truth and the denoised maps
\footnote{These maps are normalised and then Fourier-transformed.}
:
\begin{equation}
    r(\ell) = \frac{\Re \left[\langle \tilde{\kappa}_\mathrm{denoised}(\ell) \tilde{\kappa}^*_\mathrm{truth}(\ell) \rangle\right]}
    {\sqrt{\langle|\tilde{\kappa}_\mathrm{denoised}|^2 \rangle \langle|\tilde{\kappa}_{\mathrm{truth}}|^2\rangle}},
\end{equation}
where $\Re [\cdot]$ corresponds to the operation to take the real part and $\langle \cdot \rangle$
is the average over 1,000 maps.
As naively expected, both GAN and DM reconstruct the target fields at large scales to some extent,
and the performance worsens for smaller scales.
In terms of the cross-correlation coefficients,
a significant difference between GAN and DM cannot be seen.
Though it is counterintuitive, the cross-correlation with the denoised maps becomes worse
compared with that with the noisy maps. It implies that denoising fails to reproduce
the local structures of the mass maps, and this degradation may be connected
to the low performance of peak identification in the denoised maps.
That demonstrates the limitation of the current implementation of GAN- and DM-based denoising.

\subsubsection{Bispectrum}
Next, we focus on the non-Gaussian statistics, the bispectrum.
The bispectrum is the counterpart in Fourier space of the three-point correlation function
and conveys information beyond the two-point correlation functions \citep[see, e.g.,][]{Takada2004}.
The bispectrum of the convergence field $B (\bm{\ell}_1, \bm{\ell}_2, \bm{\ell}_3)$ is defined as
\begin{equation}
\langle \tilde{\kappa}(\bm{\ell}_1) \tilde{\kappa}(\bm{\ell}_2) \tilde{\kappa}(\bm{\ell}_3) \rangle = (2\pi)^2 \delta_\mathrm{D} (\bm{\ell}_1 + \bm{\ell}_2 + \bm{\ell}_3) B(\bm{\ell}_1, \bm{\ell}_2, \bm{\ell}_3).
\end{equation}
The lensing bispectrum can be expressed as the matter bispectrum
$B_\delta (\bm{k}_1, \bm{k}_2, \bm{k}_3)$ convoluted with
the lensing kernel $W_\kappa$ (Eq.~\ref{eq:kernel}):
\begin{equation}
B(\bm{\ell}_1, \bm{\ell}_2, \bm{\ell}_3) = \int_{0}^{\chi_\mathrm{H}} d\chi \, \frac{W_\kappa^3(\chi)}{r^4(\chi)} B_\delta \left( \frac{\bm{\ell}_1}{r(\chi)}, \frac{\bm{\ell}_2}{r(\chi)}, \frac{\bm{\ell}_3}{r(\chi)} \right).
\end{equation}
In this study, we employ the reduced bispectrum $\tilde{B}$
which are computed from the normalised convergence maps.
The reduced bispectrum is computed based on the estimator of \citet{Scoccimarro2015}
with the multipole range $\ell \in [200, 10000]$ with $20$ logarithmically-spaced bins.
Specifically, we consider three configurations:  
equilateral ($\ell_1 = \ell_2 = \ell_3$),
flatten ($\ell_1 = \alpha \ell_2 = \alpha \ell_3$ with $\alpha = 0.556$),
and squeezed ($\ell_1 = \ell_2 > \ell_3$ with $\ell_3 = 220.549$) configurations.

\begin{figure}
  \begin{center}
  \includegraphics[width=\columnwidth]{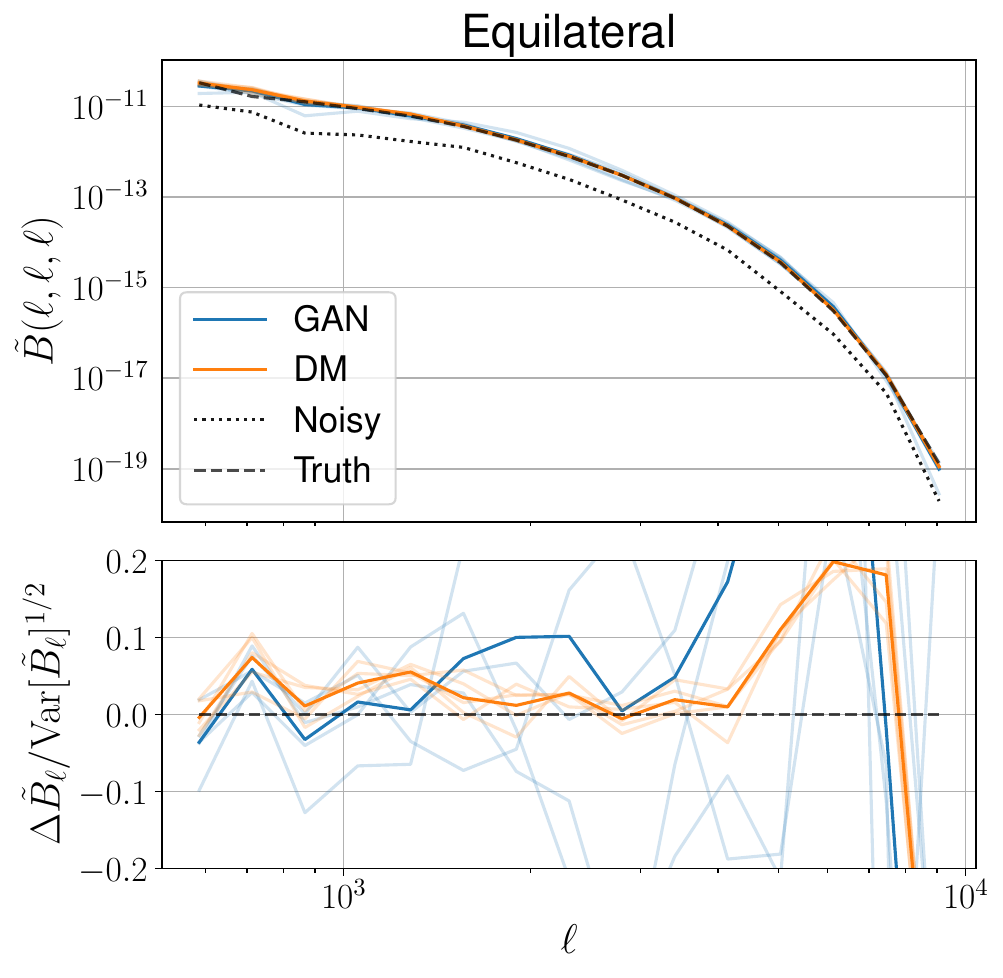}
  \end{center}
  \caption{The bispectra of the denoised maps with GAN (blue) and DM (orange)
  for equilateral configuration ($\ell_1 = \ell_2 = \ell_3 = \ell$).
  The results from the noisy maps are shown as the dotted line.
  Note that the maps are normalised by the standard deviation, and thus,
  the amplitude of bispectra from the denoised maps is higher.}
  The black dashed line corresponds to the bispectra of the true maps.
  The five blue (orange) thin lines correspond to the five networks (samples) from GAN (DM).
  The thick blue (orange) line shows the average bispectrum for GAN (DM).
  {Alt text: Three line graphs with two panels.}
  \label{fig:bl_equilateral}
\end{figure}

\begin{figure}
  \begin{center}
  \includegraphics[width=\columnwidth]{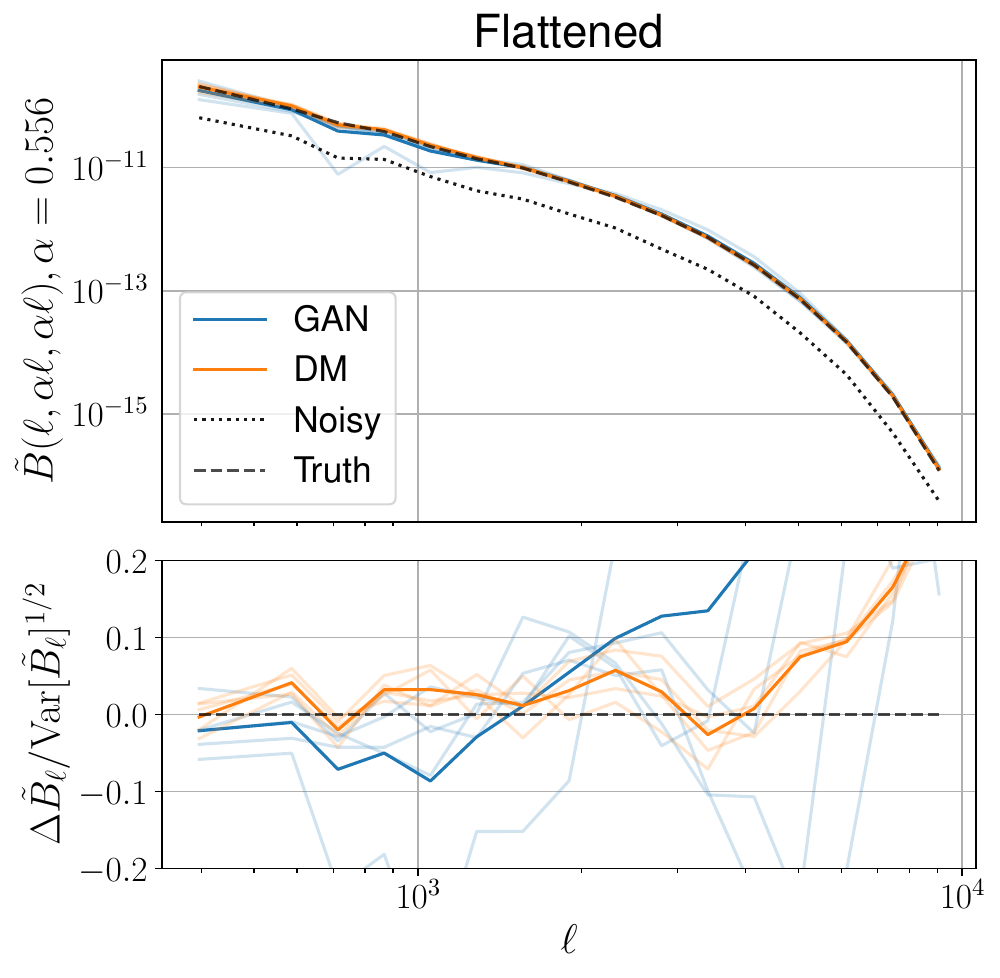}
  \end{center}
  \caption{Same as Figure~\ref{fig:bl_equilateral}, but for the bispectrum
  for flattened configuration ($\ell_1 = \ell, \ell_2 =\alpha \ell, \ell_3 = \alpha \ell$ with $\alpha = 0.556$).
  {Alt text: Three lines graphs with two panels.}}
  \label{fig:bl_flattened}
\end{figure}

\begin{figure}
  \begin{center}
  \includegraphics[width=\columnwidth]{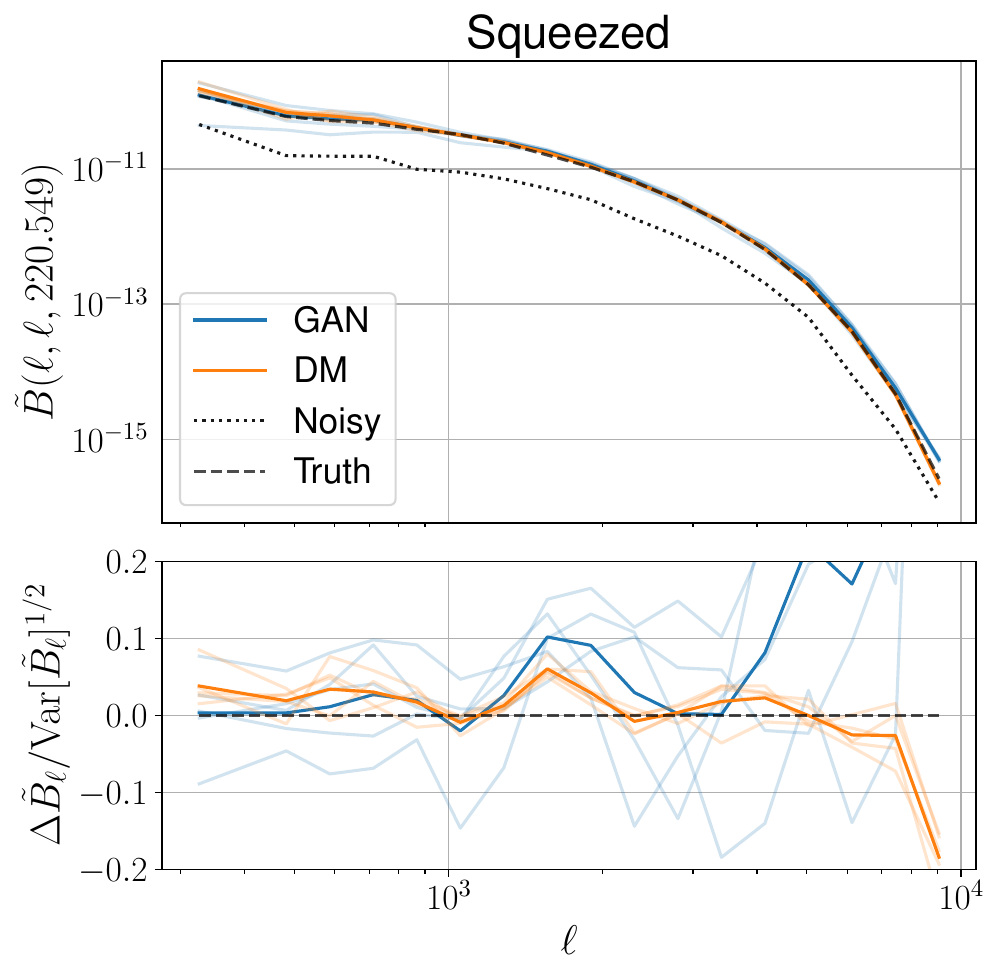}
  \end{center}
  \caption{Same as Figure~\ref{fig:bl_equilateral}, but for the bispectrum
  for squeezed configuration ($\ell_1 = \ell, \ell_2 = \ell, \ell_3 = 220.549$).
  {Alt text: Three line graphs with two panels.}}
  \label{fig:bl_squeezed}
\end{figure}

Figures~\ref{fig:bl_equilateral}, \ref{fig:bl_flattened}, and \ref{fig:bl_squeezed} show the bispectra
for the equilateral, flattened, and squeezed configurations, respectively.
DM reproduces the bispectrum for all configurations
within $0.05\sigma$ level down to small scales $ \ell \lesssim 4000$
while the accuracy of GAN exceeds $1\sigma$ at relatively large scales.
In addition, variations across different samples of DM are clearer than in the power spectrum.
For GAN, variations similar to the case in the power spectrum appear across different networks
and the fluctuation amplitude is larger than that of DM.
The features do not vary for different configurations.

\subsubsection{Probability distribution function}
\begin{figure}
  \begin{center}
  \includegraphics[width=\columnwidth]{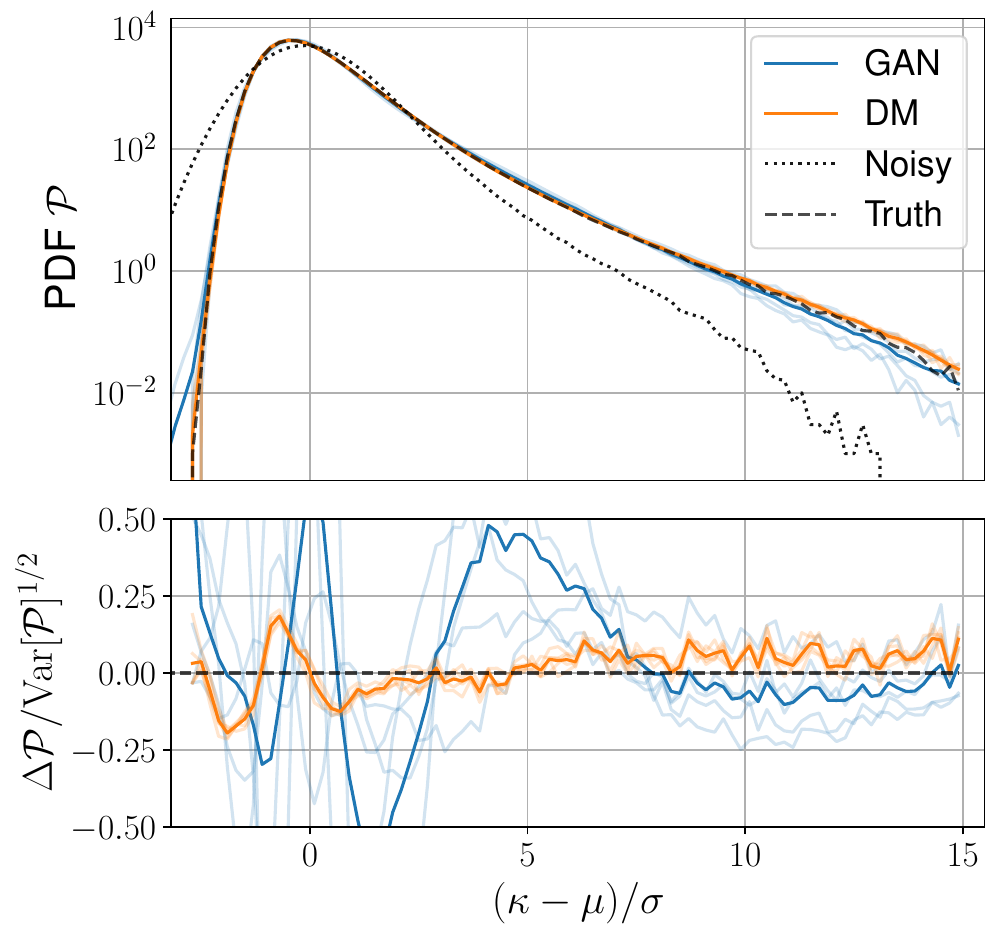}
  \end{center}
  \caption{PDFs of the denoised maps with GAN (blue) and DM (orange).
  The maps are normalised so that the PDF has zero mean and unit variance.
  The five blue (orange) thin lines correspond to the five samples from GAN (DM).
  The thick blue (orange) line shows the average PDF for GAN (DM).
  The black dashed (dotted) line corresponds to the result of true (noisy) maps.
  {Alt text: Four line graphs with two panels.}}
  \label{fig:PDF}
\end{figure}

Next, we focus on the one-point probability density function (PDF) of the denoised convergence maps.
The PDF also contains information beyond the two-point statistics and can be easily measured
in actual observations \citep{Thiele2023}.
We measure the PDF of the normalised convergence map with the range $\SNR \in [-5, 15]$
with 100 linearly spaced bins and the regions closer to the edges within $2\theta_\mathrm{G}$ are excluded.
The PDFs of the denoised maps with GAN and DM are shown in Figure~\ref{fig:PDF}.
The reconstruction accuracy of DM is at least $0.2\sigma$ for the entire range.
All five samples show a similar trend and
this consistency demonstrates the robustness of the sampling process inherent to the DM.
On the other hand, the results of GAN yield larger variability among five trained networks.
All GAN samples show deviations greater than $0.25\sigma$ in some range,
with significant deviations around $\SNR \simeq 0$ and $5$.

\subsubsection{Peak and minima counts}
Here, we discuss the peak and minima counts.
Peaks (minima) are defined as pixels with values larger (smaller) than their eight surrounding pixels,
and thus, the identification of peaks and minima in the actual analysis is straightforward.
High peaks have the clear physical origin;
massive dark matter halos are the source of the high signal \citep{Yang2011}.
The low peaks originate from the line-of-sight alignment of smaller halos.
In contrast, minima are negative peaks and mostly correspond to the underdense regions.
Therefore, peaks and minima are employed to search overdense and underdense structures in the Universe.
Furthermore, the counts of peaks and minima as a function of the height or depth can be used to constrain cosmological parameters
\citep{Coulton2020,Marques2024}.

\begin{figure*}
  \begin{center}
  \includegraphics[width=17cm]{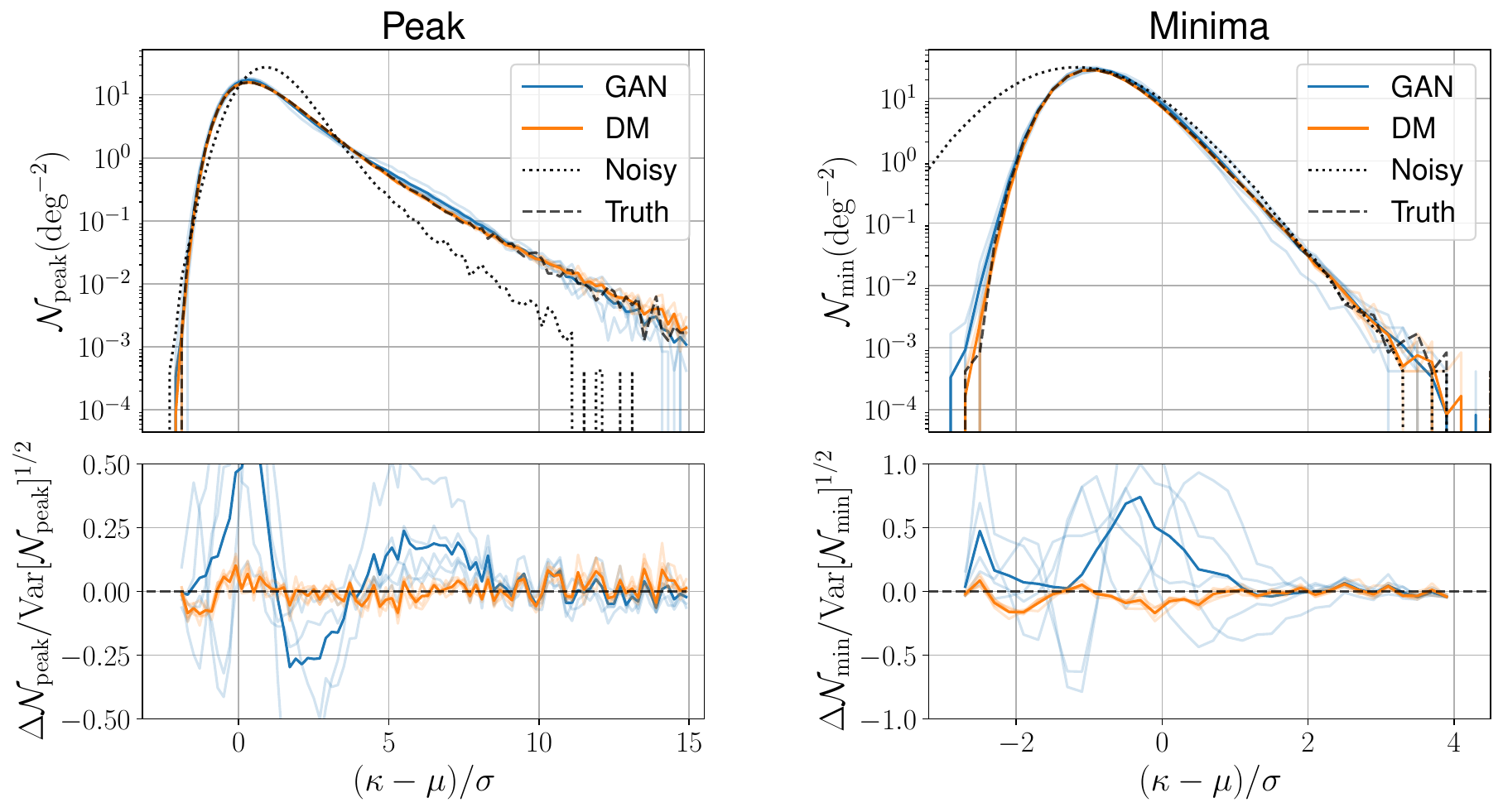}
  \end{center}
  \caption{The peak and minima counts of the denoised maps with GAN (blue) and DM (orange)
  as a function of peak height or minima depth.
  The maps are normalised so that the PDF has zero mean and unit variance.
  The five blue (orange) thin lines correspond to the five samples from GAN (DM).
  The thick blue (orange) line shows the average peak and minima counts for GAN (DM).
  The black dashed (dotted) lines correspond to the results of true (noisy) maps.
  {Alt text: Four line graphs with four panels.}}
  \label{fig:peak_minima}
\end{figure*}

In this study, the counts of peaks and minima are measured in the range $\SNR \in[-5, 15]$
with 100 linearly spaced bins and the regions closer to the edges within $2\theta_\mathrm{G}$ are excluded.
The results are presented in Figure~\ref{fig:peak_minima}.
DM can reconstruct the peak counts over the entire $\SNR$ range with an accuracy of $0.1 \sigma$.
In contrast, GAN can reproduce the peak counts only at very high peaks ($\SNR \gtrsim 8$),
and its overall performance is worse than DM. 
For the minima counts, DM can also reproduce them with an accuracy of $0.2 \sigma$ across the entire $\SNR$ range.
Each GAN sample shows deviations of roughly $0.5\sigma$ at lower depth ($\SNR \lesssim 2$) and the mean still deviates by $0.5\sigma$ around $\SNR \simeq 0$.
The reconstruction accuracy at low $\SNR$ strongly depends on the noise scheduling
and we find that the quadratic scheduling, which is our fiducial choice,
yields the best performance.
The comparison between different noise scheduling schemes is presented in
Appendix~\ref{sec:noise_schedule}.

\subsubsection{Scattering transform}
Scattering transform, originally proposed by \cite{Mallat2012},
is a method for extracting information from high-dimensional data,
and its application in cosmology has been gaining attention in recent years.
The scattering transform coefficients can capture a significant amount of non-Gaussianity in convergence maps
and have been shown to provide stronger constraints on cosmological parameters
compared to the power spectrum, peak counts, and even convolutional neural network (CNN) approach \citep{Cheng2020}.
The scattering transform applies wavelet convolutions and modulus operations to the input image,
transforming it into new representations for statistical analysis.
Due to its use of convolution and non-linear modulus operations,
it shares a structural similarity with CNNs.
While CNNs have recently been able to improve parameter constraints through advances in ML \citep{Sharma2024},
the scattering transform has the advantage of being non-trainable, making it a useful alternative.

In this study, we follow the approach presented in \citet{Cheng2020} and utilize \texttt{scattering} package\footnote{\url{https://github.com/SihaoCheng/scattering_transform}}
to compute the scattering transform coefficients.
Given an input convergence map $I_0$, we apply the first-order scattering transform to obtain the first-order representation $ I_1 $:
\begin{equation}
I_1(j_1, l_1) \equiv \left| I_0 \star \Psi^{j_1, l_1} \right|,
\end{equation}
where $\star$ denotes the convolution operation, $\Psi^{j_1, l_1}$ is the wavelet with scale index $j_1$ and orientation index $l_1$,
and $ | \cdot | $ denotes the modulus of a complex field.
As the wavelet, we employ Morlet wavelets, which are defined as
\begin{equation}
\Psi^{j,l} = \frac{1}{\sigma}\exp\left( -\frac{x^2}{2\sigma^2} \right)
\left[ e^{i\bm{k}_0\cdot\bm{x}}-\exp\left(-\frac{k_0^2\sigma^2}{2} \right) \right],
\end{equation}
where $\sigma = 0.8\times2^j$, $k_0 = \frac{3\pi}{4}\times2^{-j}$, and $\arg (\bm{k}_0) =\frac{\pi l}{L}$.  
In this study, we set $j=0, \ldots, 6$ and $L=4$, i.e., we consider a total of seven different scales and four directional Morlet wavelets.
The second-order representation is computed in a similar manner:
\begin{equation}
I_2(j_1, j_2, l_1, l_2) \equiv \left| I_1(j_1, l_1) \star \Psi^{j_2, l_2} \right|
= \left| \left| I_0 \star \Psi^{j_1, l_1} \right| \star \Psi^{j_2, l_2} \right|.
\end{equation}
\cite{Cheng2020} indicate that higher-order coefficients, i.e., third-order and beyond,
contribute to little additional cosmological information.
Therefore, we restrict our analysis to the first- and second-order scattering transform
coefficients.

From the fields $I_1$ and $I_2$ obtained via the scattering transform, we extract statistical quantities.  
Following previous works \citep{Cheng2020,Shirasaki2024}, we compute the spatial average of the field.  
Since the convergence map is isotropic, we further take an average over the directional index $l$ to compute the scattering transform coefficients $s_1(j_1)$ and $s_2(j_1, j_2)$:
\begin{align}
s_1(j_1) &= \langle I_1(j_1,l_1)\rangle_{(x,y),l_1}, \\
s_2(j_1,j_2) &= \langle I_2(j_1,j_2,l_1,l_2)\rangle_{(x,y),l_1,l_2}.
\end{align}
The region $j_2 \leq j_1$ carries little useful information for cosmological analysis.  
Therefore, we only measure the coefficients for $j_1 < j_2$.

\begin{figure*}
  \begin{center}
  \includegraphics[width=17cm]{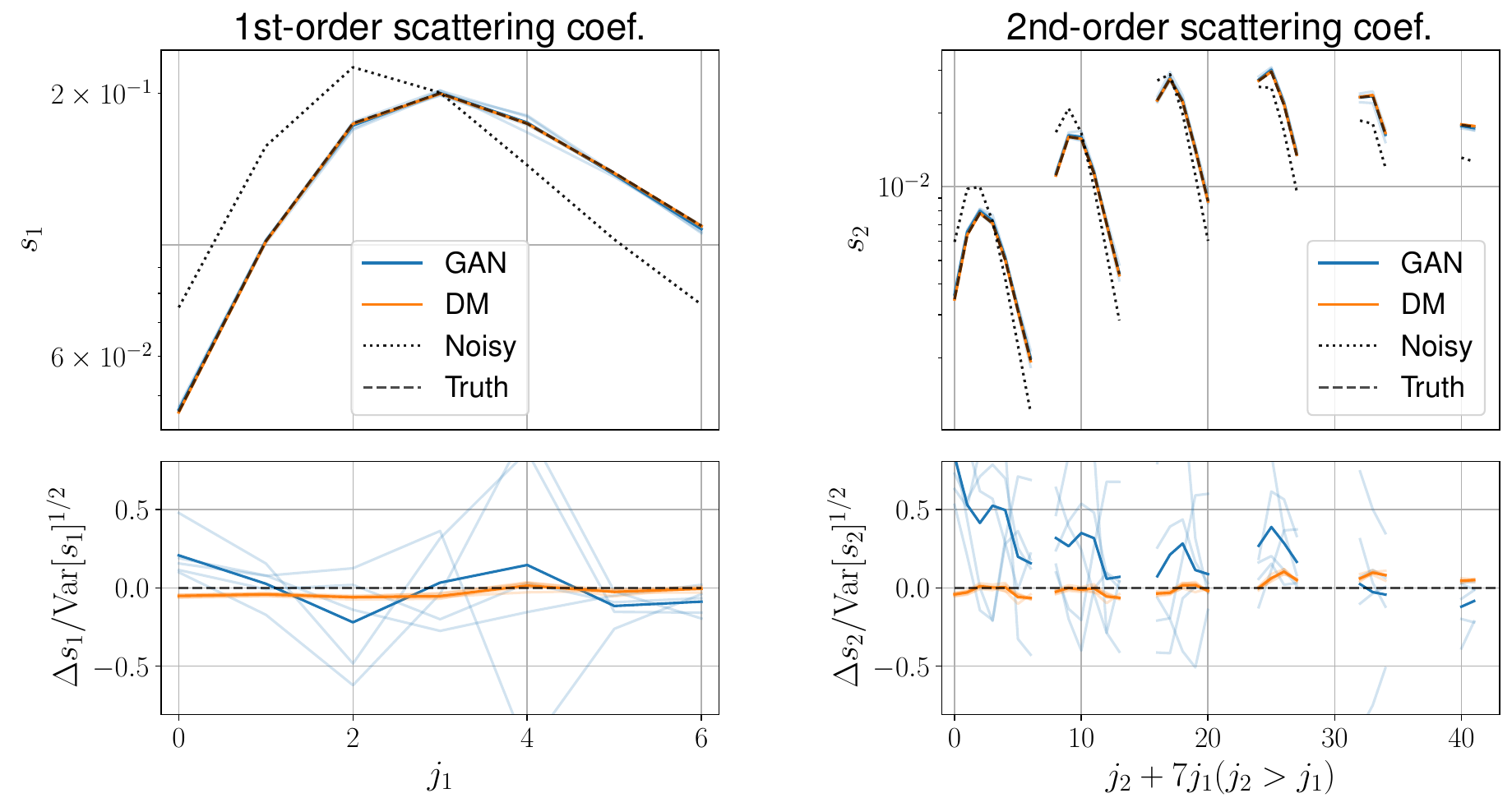}
  \end{center}
  \caption{First-order scattering coefficients $s_1(j_1)$ (left panel) and second-order scattering coefficients $s_2(j_1,j_2)$ (right panel).
  The five blue (orange) thin lines correspond to the five samples from GAN (DM).
  The thick blue (orange) line shows the average scattering coefficients for GAN (DM).
  The black dashed (dotted) lines correspond to the results of true (noisy) maps.
  {Alt text: Four line graphs with four panels.}}
  \label{fig:s_iso}
\end{figure*}

The first-order and second-order scattering transform coefficients are shown in Figure~\ref{fig:s_iso}.
Examining the first-order coefficients, we find that some GAN networks exhibit deviations that remain within $0.5\sigma$
across all scales $j_1$,
whereas all DM samples show lower deviations, i.e., below $0.1\sigma$.
The second-order coefficients show similar behaviour to the first-order ones:
DM accurately reproduces results across all scales,
and GAN exhibits significant deviations depending on the specific network. 

\subsection{Stress tests}
To evaluate the robustness and flexibility of the denoising process based on GAN and DM,
we conduct stress tests by denoising maps with characteristics different from the training data.
These stress tests mimic the real observations,
where the true properties of lensing maps, such as the source redshift, are not exactly determined.
We consider three cases: (i) maps with noise levels scaled by factors of $1.1$ and $0.9$,
(ii) maps with the convergence field multiplied by $1.1$ and $0.9$,
and (iii) maps with source galaxies at higher ($z_s = 1.2$) and lower ($z_s = 0.9$) redshifts.
These stress tests virtually correspond to applying denoising to WL surveys with
different noise characteristics (case i) or source redshift distribution (case iii),
or varying cosmological parameters (case ii).
The training is conducted only once with the fiducial data set,
and the maps in the stress tests have distinct features apart from the training data set.
For GAN, sampling is performed for five networks for each test data, which results in 5,000 denoised maps.
For DM, the variation among five samples is small, as demonstrated in Figure~\ref{fig:cl},
and thus, only one sample is employed for stress tests.
To highlight the changes relative to the fiducial test data, we show the normalised difference:
\begin{equation}
    \Delta \tilde{C}(\ell) / \mathrm{Var}[\tilde{C}(\ell)]^{1/2} \equiv
    \frac{\tilde{C}(\ell) - \tilde{C}_\mathrm{ref}(\ell)}
    {\mathrm{Var}[\tilde{C}_\mathrm{ref}(\ell)]^{1/2}} ,
    \label{eq:difference_stress}
\end{equation}
where $\tilde{C}_\mathrm{ref}(\ell)$ denotes the reference power spectrum of the ground-truth maps from modified data,
and $\tilde{C}(\ell)$ corresponds to the power spectrum of denoised maps in the stress test.
Differently from the previous results, the reference power spectrum differs according to each stress test.

\begin{figure}
  \begin{center}
  \includegraphics[width=\columnwidth]{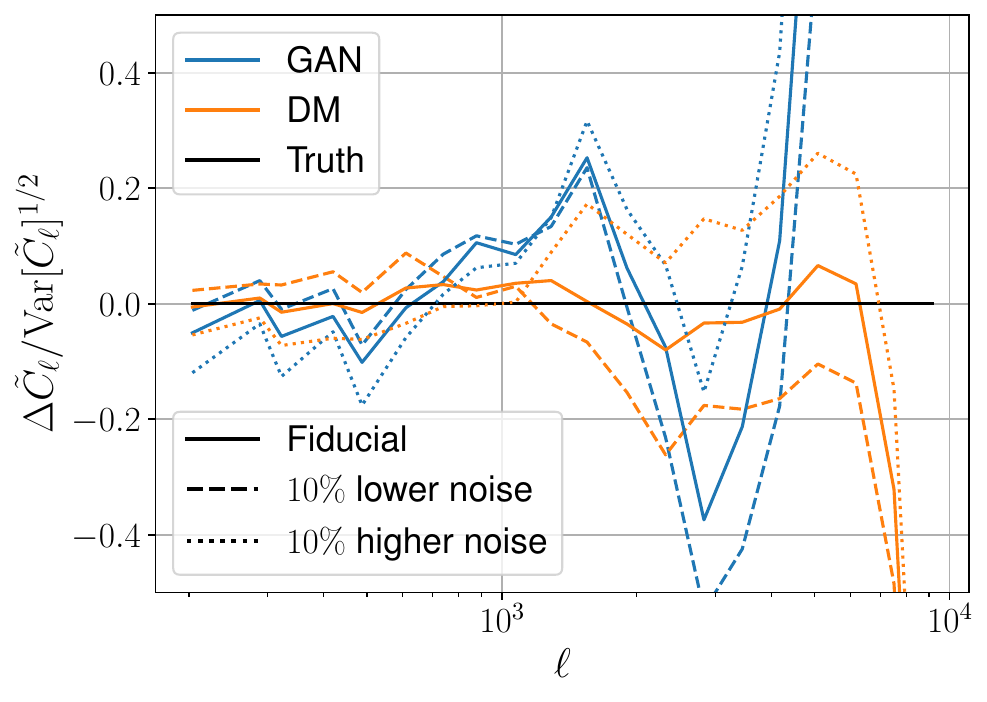}
  \end{center}
  \caption{Power spectra for stress tests with $10\%$ lower and higher noise.
  The dashed (dotted) line corresponds to the case with $10\%$ lower (higher) noise.
  Since the true convergence field is unchanged, the fractional difference for the fiducial data is always zero.
  {Alt text: Seven line graph.}}
  \label{fig:cl_stress_noise}
\end{figure}

\begin{figure}
  \begin{center}
  \includegraphics[width=\columnwidth]{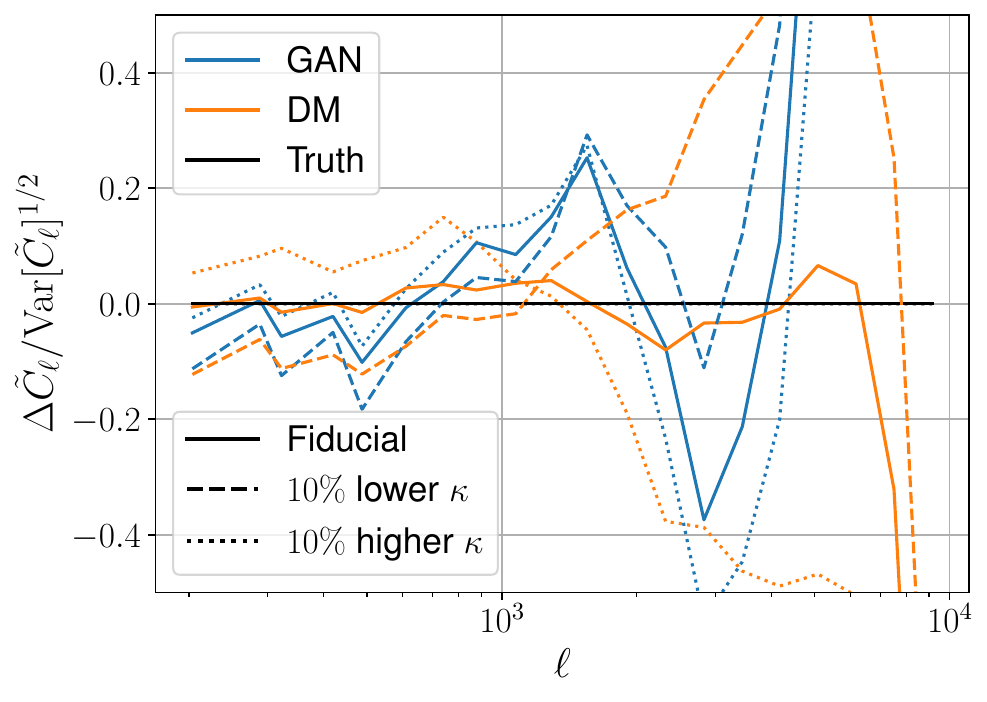}
  \end{center}
  \caption{Power spectra for stress tests with $10\%$ lower and higher convergence.
  The dashed (dotted) line corresponds to the case with $10\%$ lower (higher) convergence.
  Since only the amplitude of the convergence field is changed, the fractional difference for the fiducial data is always zero.
  {Alt text: Seven line graph.}}
  \label{fig:cl_stress_kappa}
\end{figure}

\begin{figure}
  \begin{center}
  \includegraphics[width=\columnwidth]{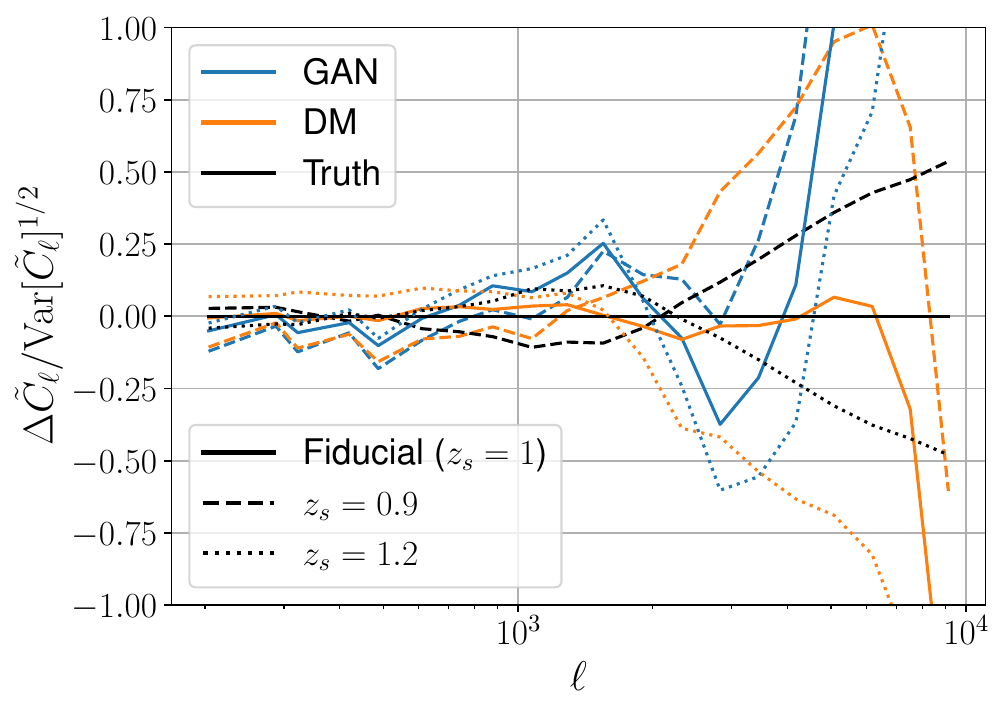}
  \end{center}
  \caption{Power spectra for stress tests with lower ($z_s = 0.9$) and higher ($z_s = 1.2$) source redshifts.
  The dashed (dotted) line corresponds to the case with lower (higher) source redshift.
  {Alt text: Nine line graph.}}
  \label{fig:cl_stress_z}
\end{figure}

In Figure~\ref{fig:cl_stress_noise}, the solid line corresponds to the power spectrum of the fiducial test data,
and the dashed (dotted) line represents the case with noise multiplied by $0.9$ ($1.1$).
The notation is the same in Figure~\ref{fig:cl_stress_kappa} for the convergence-scaled case
and in Figure~\ref{fig:cl_stress_z} for the source-shifted case.
Note that, in the noise variation case, since the convergence component remains unchanged,
the true power spectrum overlaps for the solid, dashed, and dotted lines.
The overall trend is common in all three stress tests.
The variation due to the change in the test data set is larger for DM than GAN.
However, DM still performs better in power spectrum reconstruction.

Let us investigate the details of each stress test.
When the noise amplitude is varied, the power spectra from DM denoised maps deviate
from intermediate scales ($\ell \gtrsim 1000$) and the deviation reaches $0.2 \sigma$ at $\ell \simeq 2000$,
where the noise power spectrum becomes comparable with the convergence power spectrum.
When DM is applied to the fiducial case, i.e., without noise amplitude modification,
the power spectrum is well reconstructed up to $\ell \simeq 6000$, and thus, the mismatch of the noise amplitude
leads to low performance at small scales.
A similar trend can be seen for GAN, though the reconstruction of the power spectrum fails from relatively small scales.
The discussion holds for the convergence-scaled case.
The convergence field is the target data in the denoising process,
the change in the amplitude of the convergence maps affects the reconstruction performance more;
the deviation is roughly $0.4\sigma$ at $\ell \simeq 3000$.
For the source-shifted case, we adopt $z_s = 0.9$ and $1.2$ for shifted source redshifts,
which roughly corresponds to the photometric redshift uncertainty.
The variation amplitude in power spectrum reconstruction is similar in the first and second cases.
We should note that the uncertainty of source galaxies propagates to the $0.3\sigma$ level error in power spectrum reconstruction.

\section{Conclusions}
\label{sec:conclusions}
Weak lensing (WL) is a unique probe into the large-scale structures of the Universe and has been the primary science
target in ongoing and future imaging cosmological surveys.
Through measurements of the distortion of the shapes of galaxies, the matter distribution can be mapped with WL.
The dominant contamination of the WL signal is the shape noise; the intrinsic shape of galaxies dilutes the signal
due to the finite number of source galaxies.
In this study, we apply machine learning (ML) techniques to remove the shape noise from observed noisy WL convergence maps.
This noise removal process is referred to as \textit{denoising}.
To this end, we employ image-to-image translation approaches:
generative adversarial network \citep[GAN;][]{Goodfellow2014} and diffusion model \citep[DM;][]{Sohl-Dickstein2015}.
GAN has been the major approach as the generative model, and the previous study \citep{Shirasaki2019} demonstrates
that GAN is effective at denoising WL maps.
In addition to GAN, we also investigate the denoising performance of DM, which outperforms GAN in various practical tasks
such as inpainting, super-resolution, colourisation, etc.
Furthermore, DM learns the probability distribution of denoised maps conditioned on the observed maps,
and thus, the trained model can generate multiple images from a single noisy convergence map.
This is another advantage of DM over GAN, which basically learns one-to-one mapping due to technical difficulty.

In order to evaluate the denoising performance in a realistic situation,
we make use of the large suite of mock WL maps $\kappa \mathrm{TNG}$ \citep{Osato2021}.
We construct 40,000 pairs of noisy observed mass maps and noiseless true mass maps.
Then, 39,000 maps are used to train model parameters in GAN and DM, and the remaining 1,000 maps are
used as test data.
DM can sample multiple realisations of denoised maps, and we sample five maps from each noisy map, which amounts to 5,000 denoised maps in total.
Though GAN basically can only sample a single realisation, we train five networks of GAN from different initial weights,
which mimics multiple sampling. As a result, there are also 5,000 denoised maps with GAN.
Note that the variance of 5 DM samples corresponds to the learned probability distribution,
but the variance of 5 GAN samples is the uncertainty due to failing to find the optimal network weights.

First, we investigate the pixel-level correlation between denoised maps and corresponding ground-truth maps.
Both GAN and DM can reproduce the true pixel value with statistical fluctuation to a similar extent.
Next, we study how the peaks in mass maps are affected by denoising.
In the denoised maps with GAN and DM, there are similar numbers of peaks in true maps.
However, roughly $45\%$ of peaks with $\SNR > 4$ in denoised maps are not associated with the peaks in the true map.
Moreover, $40\%$ of peaks with $\SNR > 4$ in true maps are mistakenly erased by denoising.
Therefore, our results suggest that challenges remain in accurately identifying dark matter halos and galaxy clusters when using ML-based denoising.

Next, we address how cosmological statistics can be recovered by denoising.
To begin with, we measure the angular power spectrum, which is the most fundamental statistics in WL analysis,
of denoised maps with GAN and DM. Then, we measure the higher-order statistics: bispectrum, one-point PDF, peaks and minima counts, and scattering transform.
These higher-order statistics contain information beyond the power spectrum and can be employed to place tighter constraints on cosmological parameters.
We find that DM demonstrates better performance for the reconstruction of all statistics.
In addition, the statistics of denoised maps with DM exhibit smaller fluctuations among the five samples.
On the other hand, GAN shows quite a large variance among the five networks, and thus, that leads to the caveat that
even if the mean of five GAN networks looks successful in reproducing the statistics,
the individual model may fail at reconstruction and the accuracy can degrade at $1 \sigma$ level.

Finally, we conduct stress tests to address the robustness and flexibility of the trained denoising models.
We create additional three data sets: noise scaled by $0.9$ and $1.1$, convergence scaled by $0.9$ and $1.1$,
and source redshifts shifted to $z_s = 0.9$ and $z_s = 1.2$.
Then, we denoise these maps with the model trained with original training data.
Hence, the training and test data sets are inconsistent.
However, this situation happens in real observations, where noise properties and source galaxy redshifts
are not necessarily precisely known.
We measure the power spectrum for denoised maps in the three modified test data sets.
Even if the training is conducted with different data set, denoising with GAN and DM functions to some extent, though accuracy worsens.

In this work, we explore the capabilities of GAN and DM in the specific task of denoising WL mass maps.
We reveal that in general, DM performs better than GAN, which is anticipated from the higher quality of generated images with DM in previous studies.
Moreover, GAN is known to be subject to mode collapse in training but the training of DM is more stable.
However, the computational cost of training and sampling is more demanding for DM, which hampers DM from generating a large number of samples.
There is another challenge regarding the size of mass maps.
The number of grids in mass maps is limited by computational requirements;
more grids lead to a large number of weights and may result in longer training.
In order to apply the denoising with ML-based approaches to the real observational data,
which covers wider areas than the simulation data used in this study,
some techniques such as splitting large- and small-scale modes are critical.
Furthermore, several survey realisms are simplified in the presented work,
e.g., non-uniform shape noise, survey masks, etc.
These systematics should be taken into account to assess the capabilities of
ML-based denoising approaches in realistic situations.
We leave it for future work.


\begin{ack}
This research was conducted using the FUJITSU Supercomputer PRIMEHPC FX1000 and FUJITSU Server PRIMERGY GX2570 (Wisteria/BDEC-01) at the Information Technology Center, The University of Tokyo.
\end{ack}

\section*{Funding}
This work was supported in part by JSPS KAKENHI Grant Number JP24H00215, JP25H01513, JP25K17380, JP25H00662 (K.O.) and JP24H00221 (M.S.). M.S. also aknowledge research supports by JST BOOST, Japan Grant Number JPMJBY24D8.

\section*{Data availability} 
The data underlying this article will be shared on reasonable request to the corresponding author.

\bibliographystyle{aasjournalv7}
\bibliography{main}

\appendix
\section{Architectures of networks}
\label{sec:arch_net}
Here, we describe the technical configurations of networks used in GAN and DM.

\subsection{Networks in GAN}
\label{sec:arch_net_GAN}
For GAN, the generator employs U-Net architecture \citep{Ronneberger2015},
and the discriminator utilises four convolutional blocks.
U-Net consists of eight convolutional blocks,
starting with 64 channels and increasing to 512 channels in the deepest layer.
Each convolutional block performs convolution (or transposed convolution) with a kernel size of $4 \times 4$.
The activation function used in the encoder part is LeakyReLU, in the decoder part it is ReLU,
and the tangent hyperbolic function is applied in the final layer.
Additionally, dropout with a rate of $0.5$ is applied to intermediate layers.
In the discriminator network, all convolutional layers use a $4 \times 4$ kernel, and the activation function is LeakyReLU.
The initial number of channels is 2, combining the input image and the target image.
As the layers progress, the number of channels increases up to 512, and finally, it reduces to 1 channel.
The discriminator takes only a part of the entire image to capture only high-frequency modes,
which is referred to as PatchGAN in \citet{Isola2017}. We employ $30 \times 30$ PatchGAN in the discriminator.
The batch normalisation is applied in both networks. 
The total number of parameters of networks is $5.4 \times 10^7$ in the generator
and $2.8 \times 10^6$ in the discriminator.

\subsection{Networks in DM}
\label{sec:arch_net_DM}
DM also adopts U-Net architecture for the reverse process,
and this U-Net consists of three downsampling and upsampling layers.
The input has 64 channels, and the number of channels doubles at each downsampling layer, reaching 512 at the intermediate layer.
Each layer consists of two Residual Network \citep[ResNet;][]{He2015} blocks with a kernel size of $3 \times 3$,
a self-attention layer, and a downsampling or upsampling operation.
The ResNet blocks employ group normalisation, SiLU activation function, and residual convolution.
The bottleneck processes the input sequentially through a ResNet block, a self-attention layer, and ResNet block.
The total number of parameters of the network is $6.3 \times 10^7$.

\section{Denoising performance in the high source number density case}
\label{sec:COSMOS_mock}
So far, we have assumed the source number density which mimics Subaru HSC WL survey.
We demonstrate our denoising models are effective with a higher source number density,
i.e., the case with lower shape noise amplitude.
To this end, we repeat our denoising process with higher source number density.
Here, the source number density is adopted as $n_\mathrm{g} = 64.2\,\mathrm{arcmin}^{-2}$,
which is determined based on COSMOS catalogues \citep{Schrabback2010}.
We train and test both GAN and DM in the same configuration as our fiducial mock maps,
i.e., $39,000$ maps for training and $1,000$ maps for testing,
but GAN is trained only once instead of five in the fiducial model.
The results of RMSE and Pearson correlation coefficient $\rho$ are summarised in Table~\ref{tab:RMSE_Pearson_COSMOS}.
In comparison with the fiducial result shown in Table~\ref{tab:RMSE_Pearson},
the similar level of denoising performance is confirmed with the higher source number density.

\begin{table}
    \tbl{
    RMSE and Pearson correlation coefficient $\rho$ with the high source number density data set.
    The boldface denotes the best value among the individual, mean, and median maps.
    }
    {
    \begin{tabular}{l|c|c}
        \hline
        Method & RMSE ($\times 10^{-2}$)$\downarrow$ & Pearson coef. $\rho$ $\uparrow$ \\
        \hline
        \hline
        GAN & 0.79 & 0.81 \\
        DM mean  & \textbf{0.61} & \textbf{0.88} \\
        \hline
        \hline
        No denoising & 0.82 & 0.85 \\
        \hline
    \end{tabular}
    }
    \label{tab:RMSE_Pearson_COSMOS}
\end{table}

\section{Noise schedule of diffusion model}
\label{sec:noise_schedule}
\begin{figure}
  \begin{center}
  \includegraphics[width=\columnwidth]{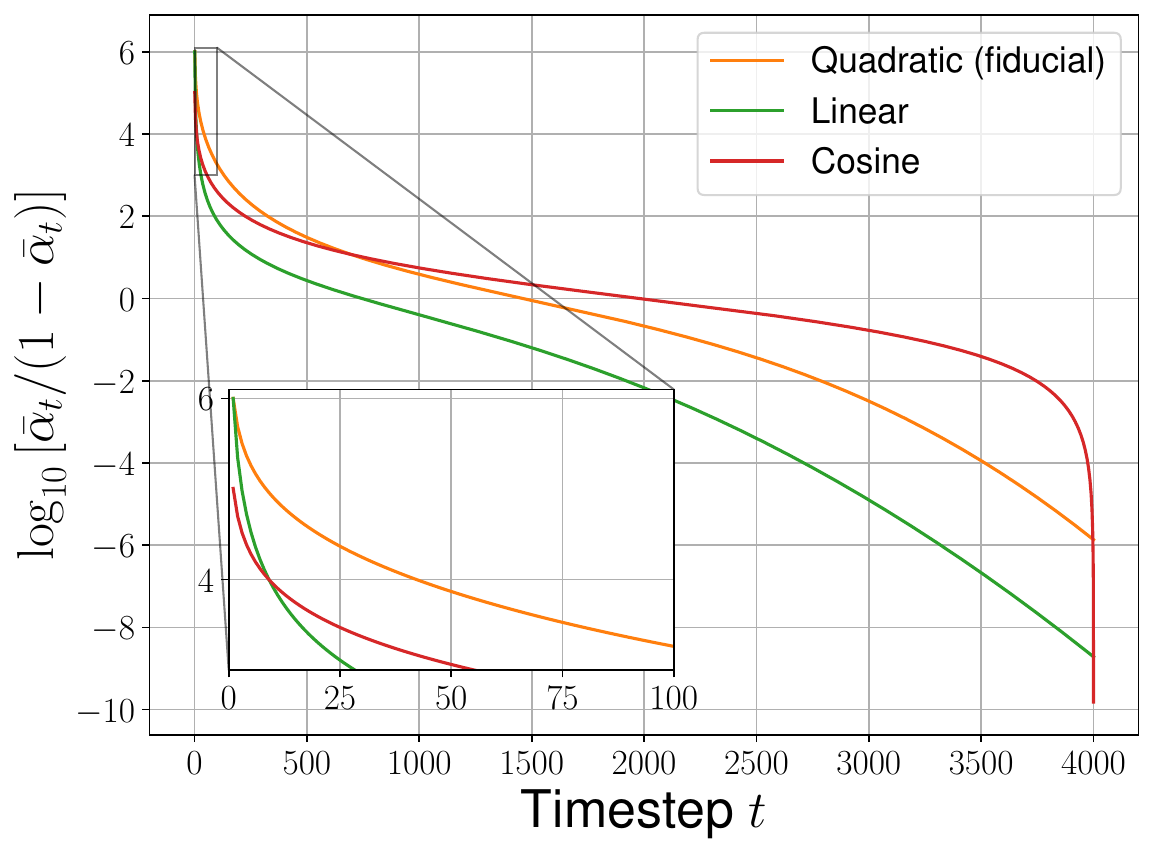}
  \end{center}
  \caption{The comparison of three noise schedules: quadratic (orange), linear (green)
  and cosine (red) schedules.
  {Alt text: Three line graphs with two panels.}}
  \label{fig:noise_schedule_comparison}
\end{figure}

Here, we address how the noise schedule $\alpha_t (=1-\beta_t)$ (Eq.~\ref{eq:diffusion})
affects the reconstructed cosmological statistics.
Three different noise schedules are considered:
linear schedule, quadratic schedule, and cosine schedule.
The linear and quadratic schedules,
which are denoted as $\beta^\mathrm{linear}_t$ and $\beta^\mathrm{quad}_t$, respectively,
start from $\beta_1 = 10^{-6}$ to $\beta_T = 0.01$
are defined as
\begin{align}
    \beta^\mathrm{linear}_t &= \beta_1  (1 - t) + \beta_T t,\\
    \beta^\mathrm{quad}_t &= \left(\sqrt{\beta_1} (1 - t) + \sqrt{\beta_T} t\right)^2.
\end{align}
The cosine schedule, which is originally proposed by \cite{Nichol2021},
is designed to prevent the diffusion process from being noise-dominated within the first few steps, which allows for more efficient training.
The cosine schedule is expressed as
\begin{align}
    \beta^\mathrm{cos}_t &= \min\left(1 - \frac{\bar{\alpha}^\mathrm{cos}_t}{\bar{\alpha}^\mathrm{cos}_{t-1}}, 0.999\right),\\
    \bar{\alpha}^\mathrm{cos}_t &= \frac{f(t)}{f(0)},\\
    f(t) &= \cos\left(\frac{\pi}{2} \frac{t/T+s}{1+s} \right)^2,
\end{align}
where $s = 8\times 10^{-3}$ is a parameter introduced to prevent $\beta_t$ from becoming too small.
The comparison of the three noise schedules is shown in Figure~\ref{fig:noise_schedule_comparison}.

\begin{figure}
  \begin{center}
  \includegraphics[width=\columnwidth]{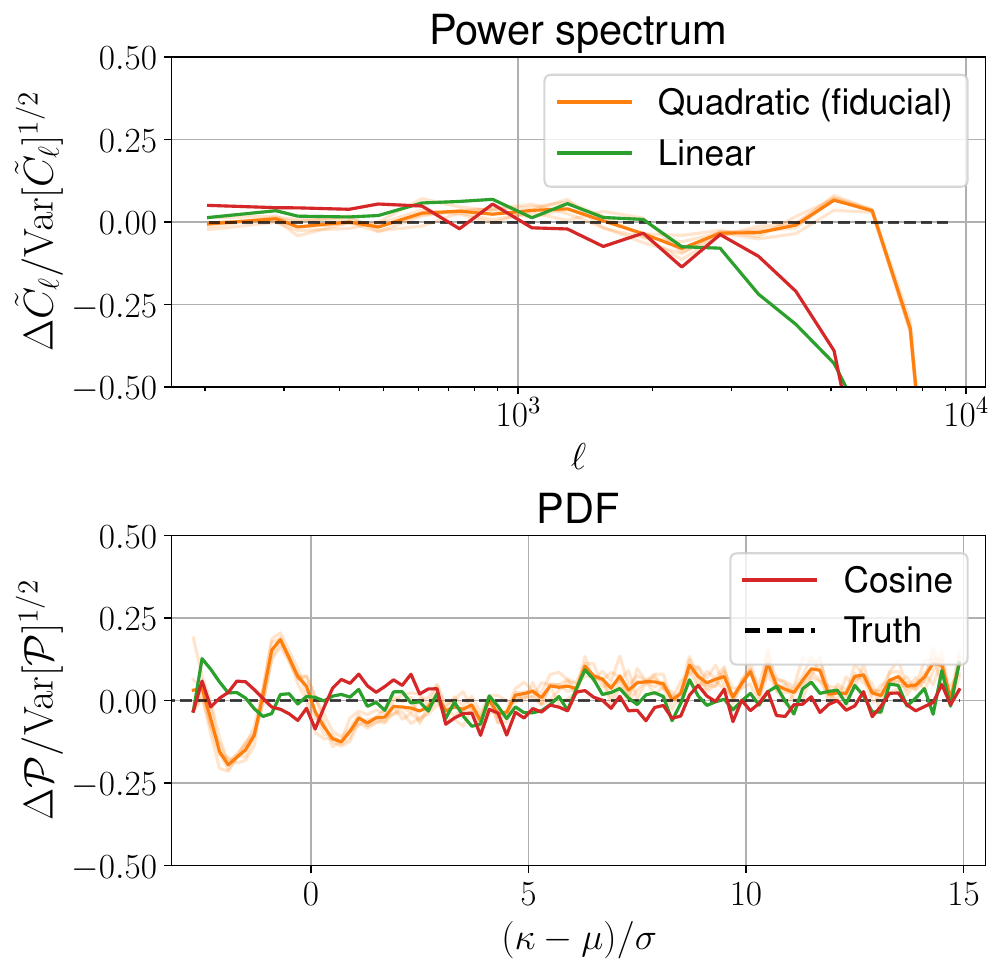}
  \end{center}
  \caption{The comparison of the power spectra and the PDF of denoised maps with DM between the three noise schedules: quadratic (orange), linear (green) and cosine (red). The solid lines of the linear and cosine schedules are obtained by a single sampling, while the quadratic schedule curves are the same as those in Figure~\ref{fig:cl} and \ref{fig:PDF}.
  {Alt text: Four line graphs with two panels.}}
  \label{fig:noise_schedule_cl_pdf}
\end{figure}

\begin{figure}
  \begin{center}
  \includegraphics[width=\columnwidth]{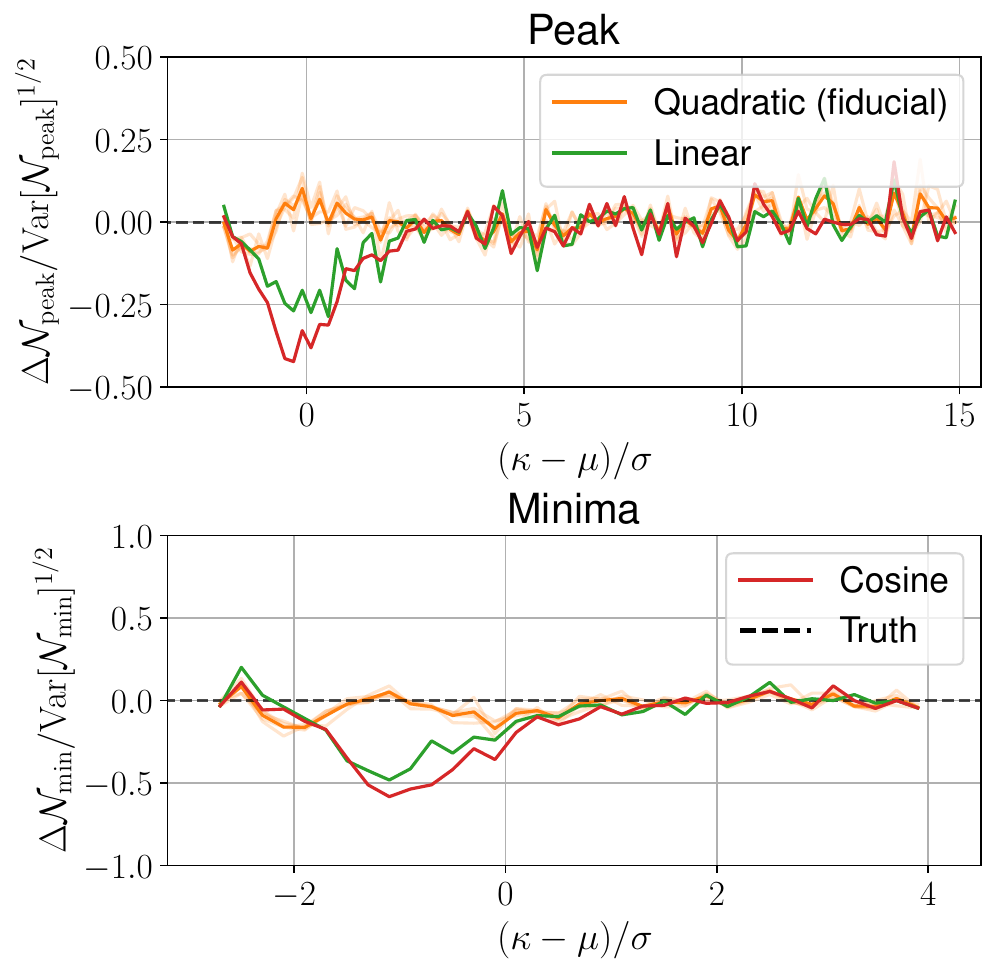}
  \end{center}
  \caption{Same as Figure~\ref{fig:noise_schedule_cl_pdf} but for peak and minima counts.
  {Alt text: Four line graphs with two panels.}}
  \label{fig:noise_schedule_peak_minima}
\end{figure}

The reconstructed power spectra, PDFs, peak and minima counts with three noise schedules are shown
in Figures~\ref{fig:noise_schedule_cl_pdf} and \ref{fig:noise_schedule_peak_minima},
Except PDFs, where the linear and cosine schedules yield slightly better reconstruction performance than the quadratic schedule,
the quadratic noise schedule reproduces the other statistics more accurately,
especially at small scales and in low-$\SNR$ regions.
This feature can be explained as the quadratic schedule takes more steps with smaller noise amplitudes
and high-frequency modes can be more efficiently learned.

\section{Direct and indirect denoising}
\label{sec:indirect}
\begin{figure}
  \begin{center}
  \includegraphics[width=\columnwidth]{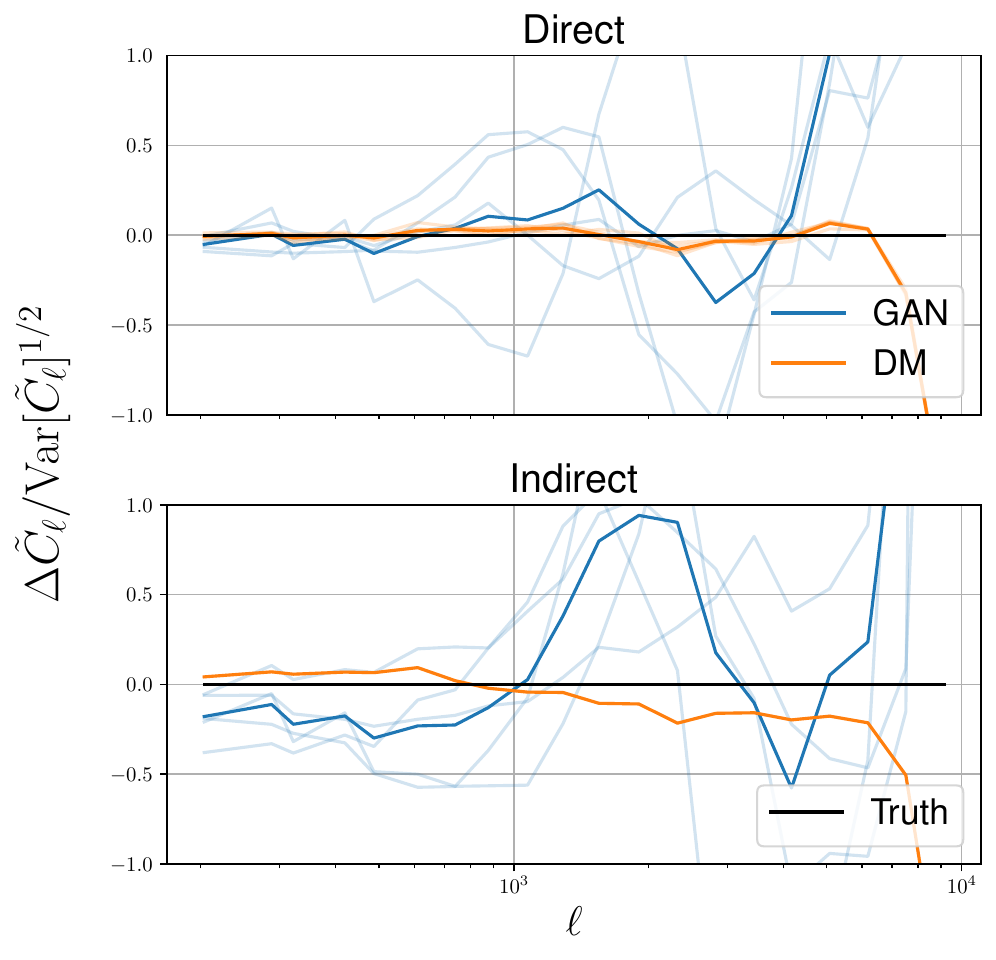}
  \end{center}
  \caption{The comparison of the power spectra of denoised maps with GAN and DM between the direct and indirect transformation.
  The upper panel shows the fractional difference from the true power spectrum when the model directly learns the noiseless convergence maps
  as the target field.
  The lower panel shows the results when the models learn the noise map, and then the noiseless convergence maps
  are obtained by subtracting the noise by the model from the noisy map.
  {Alt text: Three line graphs with two panels.}}
  \label{fig:cl_output}
\end{figure}

\begin{figure}
  \begin{center}
  \includegraphics[width=\columnwidth]{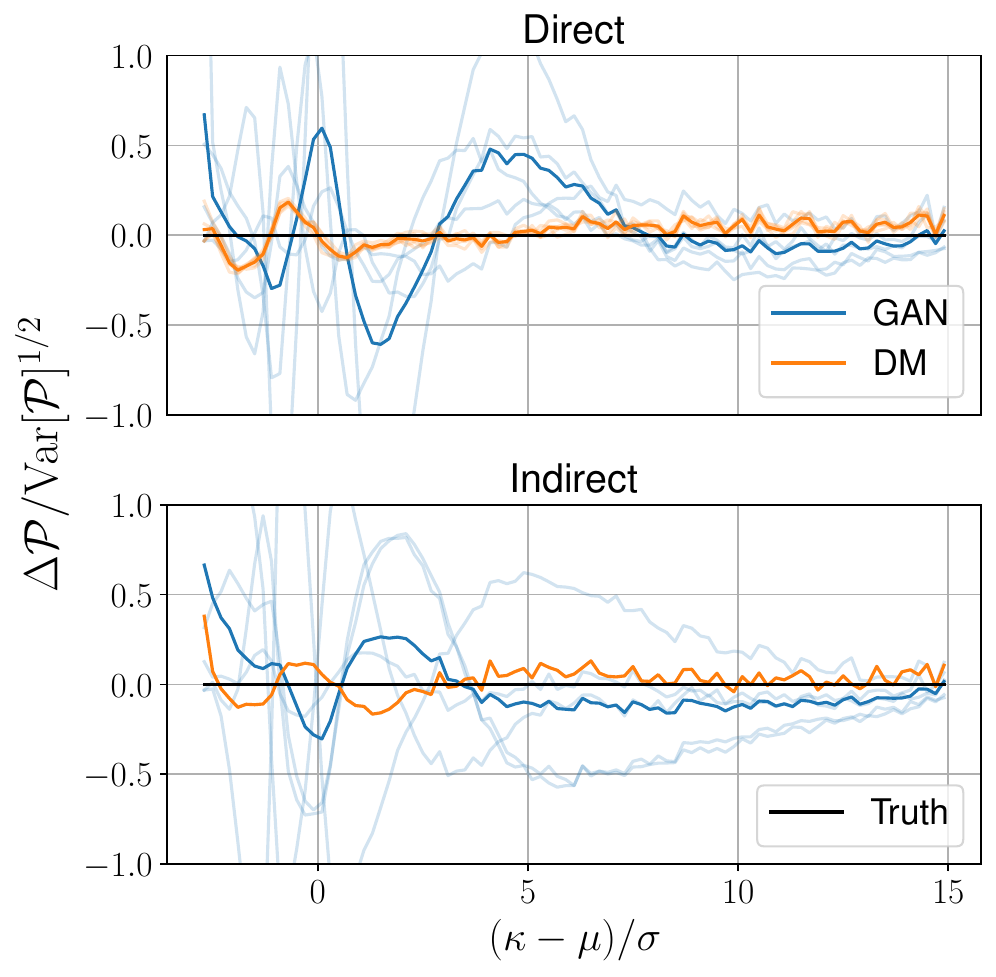}
  \end{center}
  \caption{Same as Figure~\ref{fig:cl_output} but for PDFs.
  {Alt text: Three line graphs with two panels.}}
  \label{fig:pdf_output}
\end{figure}

Here, we present the results of the comparison between the target data in the denoising process.
We define direct denoising as the case when the target field of GAN and DM is the noiseless convergence map.
On the other hand, in indirect denoising, the models learn the mapping from the noisy map,
i.e., the sum of convergence and noise maps, to the noise map itself.
Then, the noiseless mass map can be estimated by subtracting the predicted noise map from the noisy map.
The power spectra and PDFs are shown in Figures~\ref{fig:cl_output} and \ref{fig:pdf_output}, respectively
For indirect denoising, we use the same hyper-parameters as the direct denoising.
In general, it turns out that direct denoising performs better than indirect denoising.
We conclude that directly estimating the noiseless convergence map yields better results.
However, \cite{Shirasaki2019} have reported cases where the indirect transformation achieves higher accuracy,
suggesting that the optimal approach may depend on the data set.

\end{document}